\documentclass[sigconf, nonacm]{acmart}
\usepackage[linesnumbered]{algorithm2e}
\usepackage[format=plain, labelfont={bf}, textfont=normal]{caption}
\usepackage{booktabs,xcolor}
\usepackage{natbib}
\setcitestyle{authoryear}
\SetAlFnt{\small\sffamily}

\begin{document}
\title{GPU-friendly Stroke Expansion}
\author{Raph Levien}
\author{Arman Uguray}
\affiliation{
    \institution{Google}
    \city{San Francisco}
    \state{CA}
    \country{USA}
}




\begin{abstract}
    Vector graphics includes both filled and stroked paths as the main primitives. While there are many techniques for rendering filled paths on GPU, stroked paths have proved more elusive. This paper presents a technique for performing stroke expansion, namely the generation of the outline representing the stroke of the given input path. Stroke expansion is a global problem, with challenging constraints on continuity and correctness. Nonetheless, we implement it using a fully parallel algorithm suitable for execution in a GPU compute shader, with minimal preprocessing. The output of our method can be either line or circular arc segments, both of which are well suited to GPU rendering, and the number of segments is minimal. We introduce several novel techniques, including an encoding of vector graphics primitives suitable for parallel processing, and an Euler spiral based method for computing approximations to parallel curves and evolutes.
\end{abstract}


\maketitle
\thispagestyle{empty}
\pagestyle{plain}

\section{Introduction}

\begin{figure}
    \vspace{15pt}
    \includegraphics[scale=0.4]{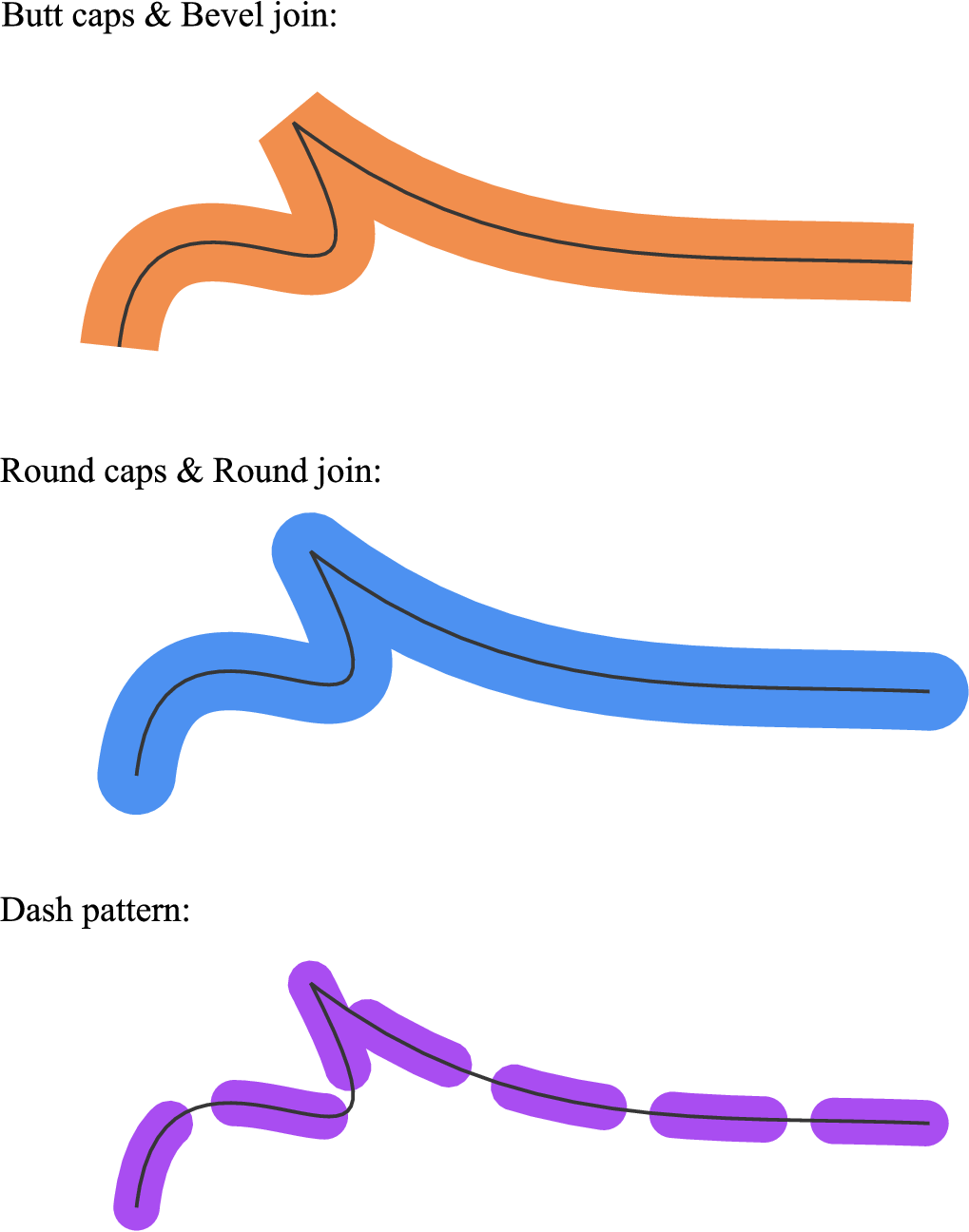}
    \caption{Illustration of stroke styles applied to a source Bézier path (shown in black).}
    \Description{Illustration of stroke styles applied to a source Bézier path (shown in black).}
    \label{fig:stroke_styles}
\end{figure}

Stroke rendering is an essential part of the standard vector graphics imaging model. The standard representation of paths is a sequence of cubic Bézier segments. While there are a number of published techniques for GPU rendering of filled paths, stroke rendering is more challenging to parallelize, largely because path segments cannot be processed independently of each other; rendering of the \emph{joins} between path segments depends on both adjoining segments, and the endpoints of paths (when not closed) are rendered with \emph{caps}. Joins and caps can be rendered with a variety of styles within the same document. The style of stroke rendering also optionally includes \emph{dashing}, which applies a pattern of dashes and gaps to the outline of the shape, as illustrated in Figure~\ref{fig:stroke_styles}.

A typical vector graphics document may have thousands of stroked paths. A map or CAD drawing may have tens of thousands of paths, possibly consisting of millions of path segments. At this scale, the computational cost of stroke rendering on a CPU may result in loss of interactive rendering performance. We believe the runtime performance can be improved significantly by offloading stroke rendering to a GPU. This requires that the algorithm be GPU-friendly: that it exploit the available parallelism, be work efficient, and avoid unnecessarily complex and divergent control flow. It must also remain numerically robust when computed with 32-bit floating pointing numbers.

There are a number of strategies for rendering strokes. Among them are \emph{local} techniques that break down the stroke into individual closed primitives, distance field techniques (or similar techniques based on point inclusion). In this paper, we focus on \emph{stroke expansion,} the generation of an outline that, when filled, represents the rendered stroke. Implementing stroke rendering using stroke expansion enables a unified approach to rendering both filled and stroked paths downstream. Among other benefits, such an approach avoids a large number of draw calls when stroked and filled paths are finely interleaved, as is often the case.

\begin{figure}
    \includegraphics[scale=0.5]{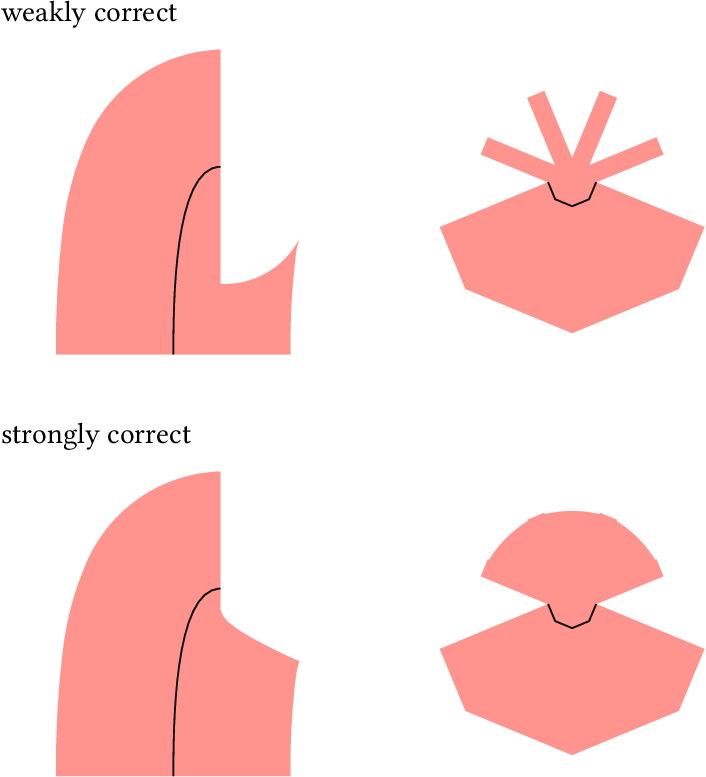}
    \caption{Example of weakly correct (top) and strongly correct (bottom) stroke rendering. The left example has a curvature cusp and requires evolutes to render correctly, while the right requires the inner join.}
    \Description{Two examples of strokes. The top left is missing an evolute section, making it appear that a half-disc has been carved out. The top right is missing inner joins, looking like spokes rather than a connected path. The bottom two examples are rendered with the full strokes.}
    \label{fig:weak_strong}
\end{figure}

Correctness is another challenging aspect of stroke rendering. We propose that the correctness of stroke outlines be divided into \emph{weak correctness} and \emph{strong correctness.} We define strong correctness as the computation of the outline of a line swept along the segment, maintaining normal orientation, combined with stroke caps and joins. Weak correctness, by contrast, only requires the parallel curves (also known as ``offset curves'') of path segments, combined with caps and the outer contours of joins. Both \citet{Nehab2020} and \citet{Kilgard2020} provide useful definitions of strongly correct stroke rendering, and Nehab in particular describes how to achieve it in stroke expansion. Briefly, when the path curvature exceeds the reciprocal of the stroke half-width, \emph{evolute} segments are required in addition to the parallel curves, as well as \emph{inner contours of joins} when segments are short. The technique described in this paper produces strongly correct outlines according to this definition. Examples of weakly and strongly correct rendering are shown in Figure~\ref{fig:weak_strong}.

Correct stroke rendering also relies on accurate approximation of the parallel curves of the input paths. Determining the parallel curve of a cubic Bézier is not analytically tractable, as it is a tenth order algebraic formula~[\citenum{Farouki1990}]. Thus, all practical stroke rendering techniques employ an approximation, with error tolerance as a tunable parameter, typically representing the maximum allowable Fréchet distance between the true curve and its approximation. A typical value is 0.25 device pixel, as that is close to the threshold of visibility. A correct algorithm will produce an approximate curve within the error tolerance, but an efficient algorithm will not subdivide significantly more than is necessary to meet the error bound. Some specifications for vector graphics, including PDF~[\citenum{PDF2008}], specify the error tolerance explicitly.

The subproblem of approximating parallel curves has spawned extensive literature, none of which is entirely satisfying, especially when it comes to algorithms that can be efficiently evaluated on GPU. An example of a reasonably good algorithm that computes flattened parallel curves is \citet{Yzerman2020}, as used in the Blend2D rendering library. For producing curved outlines, the Tiller and Hanson~[\citenum{Tiller1984}] algorithm is commonly implemented and cited, but its performance is quite poor when applied to cubic Béziers. The quadratic Bézier version is adequate, and forms the basis of the algorithm in \citet{Nehab2020}. A variant is also used in Skia~[\citenum{Skia}]; instead of lowering the cubic Bézier to a quadratic approximation and then computing the parallel curve, Skia approximates the parallel curve directly with a quadratic Bézier, then measures the error to determine whether further subdivision is necessary.

Previous work on stroke rendering does not fully solve the problem of fast and correct stroking. Most existing implementations have correctness problems, in particular failing to draw the evolutes and inner joins as required for strong correctness; \citet{Nehab2020} provides a detailed survey. In addition, most implementations are not suitable for running on GPU, requiring sequential execution to produce watertight outlines. One GPU implementation is polar stroking (\citet{Kilgard2020}), a local technique, but among other limitations, it does not bound the Fréchet distance error (rather, it bounds the angle error).

Our central approach bounding the error tolerance is \emph{error metrics,} which predict the Fréchet distance error. The error metrics presented in this paper are both straightforward to compute, and also \emph{invertible}, which avoids excessive subdivision and also increases the amount of available parallelism to exploit. Previous approaches to error measurement generally follow a ``cut then measure'' approach, where the approximate curve is generated, then the error measured by sampling. Such sampling is often time consuming, involving iteration, and even then risks underestimating the error due to inadequate sampling.

Previous work, particularly \citet{Kilgard2020}, has explained how to break the input path into dash segments of the specified length on GPU using a prefix sum of segment arc lengths. There are implementation tradeoffs to GPU-based dash processing, including potentially a larger number of dispatch stages. We anticipate that the potential performance benefit of implementing dashing on a GPU depends on the scene and other factors in the overall rendering pipeline. While we would expect further performance improvement from GPU-based dashing, we did not fully explore the solution space and we were able to achieve acceptable performance with dash processing on CPU.

This paper presents a solution to the core problem of stroke expansion, well suited to GPU implementation. It can produce both flattened polylines or an approximation consisting of circular arcs. The best choice depends on the capabilities of the path rendering mechanism following stroke expansion, but as we show, outlines consisting of circular arcs have many fewer segments and are correspondingly faster to produce. The algorithm is based on Euler spiral segments as an intermediate representation, with an iterative algorithm based on a straightforward error metric for conversion from cubic Béziers. We have also devised a compact binary encoding of paths, suitable for fully parallel computation of stroke outlines, while requiring minimal CPU-side processing. Our algorithm has been implemented in GPU compute shaders, integrated in a full rendering engine for vector graphics, and shows a dramatic speedup over CPU methods.

\section{Flattening and arc approximation of curves}


The core problem in stroke expansion is approximating the desired curve by segments of some other curve, usually a simpler one. These segments must be within an error tolerance of the source curve, and ideally close to a minimal number of them. We consider a number of source-to-target pairs, most importantly cubic Béziers to Euler spirals, and Euler spiral parallel curves to either lines or arcs.

There are generally three approaches to such curve approximation. The most straightforward but also least efficient is ``cut then measure,'' usually combined with adaptive subdivision. In this technique, a candidate approximate curve is produced, then the error is measured against the source curve, usually by sampling a number of points along both curves and determining a maximum error (or perhaps some other error norm). If the error is within tolerance, the approximation is accepted. Otherwise, the curve is subdivided (usually at $t = 0.5$) and each half is recursively approximated. A substantial fraction of all curve approximation methods in the literature are of this form, including \citet{Nehab2020}. The main disadvantage is the cost of computing the error metric. Another risk is underestimating the error due to inadequate sampling; this is a particular problem when the source curve contains a cusp.


The next approach is similar, but uses an \emph{error metric} to estimate the error. Ideally such a metric is a closed-form computation rather than requiring iteration. A good error metric is conservative, yet tight, in that it never underestimates the error (which would allow results exceeding the error bound to slip through), and does not significantly overestimate the error, which would result in more subdivision than optimal.

By far the most efficient approach is an \emph{invertible} error metric. In this approach, the error metric has an analytic inverse, or at least a good numerical approximation. Because the metric is invertible, it can predict the number of subdivisions needed, as well as the parameter value for each subdivision. If the error metric is accurate, then approximation is near-optimal. One example of an invertible error metric is angle step, used in polar stroking (\citet{Kilgard2020}); the number of subdivisions is the total angle subsumed by the curve divided by the angle step size, and the parameter value for each subdivision is the result of solving for a tangent direction. Another widely used invertible error metric is Wang's formula (\citet{Goldman2003}, Section 5.6.3), which gives a bound on the flattening error based on the second derivative of the curve. This metric is conservative but works well in practice; among other applications, it is used in Skia for path flattening. The chief limitation of Wang's method is that it computes a subdivision bound for a Bézier curve without accounting for the displacement of the parallel curve due to stroking. When applied naively to generation of parallel curves, it can undershoot substantially, especially near cusps.


\subsection{Error metrics for flattening}

The distance between a circular arc segment of length $s$ and its chord, with angle between arc and chord of $\theta$ (see Figure~\ref{fig:arc_chord}), is exactly $(1-\cos \theta)\frac{s}{2\theta}$. The curvature is $\kappa = \frac{2\theta}{s}$ (equivalently, $\theta = \frac{\kappa s}{2}$), and this remains constant even as the arc is subdivided. Rewriting, $d = (1-\cos\frac{\kappa s}{2})\frac{1}{\kappa}$. From this, we can derive a precise, invertible error metric. Subdividing the arc into $n$ segments, the distance error for each segment is $\frac{1}{\kappa}(1-\cos \frac{\kappa s}{2n})$. Solving for $n$, we get:

\begin{figure}
    \includegraphics[scale=0.5]{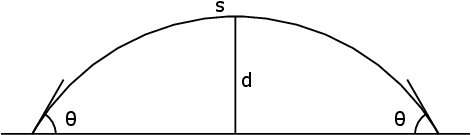}
    \caption{A circular arc segment with notations for angles ($\theta$), arc length ($s$), and distance to chord ($d$)}
    \Description{A circular arc above its chord, with labels at the endpoints for angle from chord ($\theta$), the length of the arc ($s$) and the maximum distance from the arc to its chord ($d$).}
    \label{fig:arc_chord}
\end{figure}


\[
n = \frac{s\kappa}{2\cos^{-1}\left( 1- d\kappa \right)}
\]

To flatten a finite arc, round up $n$ to the nearest integer. This will cause the error to decrease, so will still be within the error bounds.

Note that the number of subdivisions is proportional to the arc length. Another way of stating this relationship is that the \emph{subdivision density}, the number of subdivisions per unit of arc length, is constant.

The error metric for flattening an arc is exact. It always yields the minimum number of subdivisions needed to flatten the curve, and the flattening error is the least possible given that number of subdivisions. For general curves, an exact error bound is not feasible, and we resort to an approximation. Again the circular arc provides a good example. Applying the small angle approximation $\cos \theta \approx 1 - \theta^2/2$, the approximate distance error is $d = \frac{\kappa s^2}{8n^2}$, and solving for $n$ we get $n = s\sqrt{\frac{\kappa}{8d}}$. Note that this estimate is \emph{conservative,} in that it will always request more subdivision and thus produce a lower error than the exact metric.

We are of course concerned with the flattening of more general curves (ultimately the parallel curve of a cubic Bézier), not simply circular arcs, so the curvature is not constant. An obvious approach to try would be to use the maximum error. This would be conservative with respect to error tolerance, but also generate too much subdivision when the curve has a small region of high curvature. In the limit, the curve can contain a cusp of infinite curvature, which would entail infinite subdivision. Similarly, sampling the curvature at a single point can either underestimate or overestimate the error. The ideal approach would be a form of average curvature that is tractable to compute, and accurately predicts the flattening error. To that end, we propose the following error metric to estimate the maximum distance between an arbitrary curve of length $\hat{s}$ and its chord.

\[
    d \approx \frac{1}{8}\left(\int_0^{\hat{s}} \sqrt{|\kappa(s)|}ds \right)^2
\]


This formula is the same as the approximate error metric for circular arcs, except that instead of a constant curvature value, a norm-like average with an exponent of 1/2 is used (it is not considered a true norm because the triangle inequality does not hold). This particular formulation has two strong advantages. First, it has been empirically validated to accurately predict the distance between chord and curve for a variety of curves. In particular, we believe that when curvature is monotonic, it is a tight bound both above and below (and later we will see that in our algorithm it is always evaluated on curve segments with such monotonic curvature). Second, this formula lends itself to invertible error metrics.

This formula also has a meaningful interpretation: the quantity under the integral sign is the subdivision density, and represents the number of subdivisions per unit length of an optimal flattening as the error tolerance approaches zero. In particular, the number of subdivisions is:

\[
    n = \left(\int_0^{\hat{s}} \sqrt{|\kappa(s)|}ds \right)\sqrt{\frac{1}{8d}}
\]

In addition, if the function represented by the integral is invertible, then the corresponding error metric is invertible. Evaluate the function to determine the number of subdivision points, then evenly divide the result, using the inverse of the function to map these values back into parameter values for the source curve being approximated.

\subsection{Error metrics for flattening Euler spirals}

We choose Euler spiral segments for our intermediate curve representation precisely because their simple formulation in terms of curvature (Cesàro equation) results in similarly simple subdivision density integrals.

An Euler spiral segment is defined by $\kappa(s) = as+b$, or alternatively $\kappa(s) = a(s-s_0)$, where $s_0 = -b/a$ is the location of the inflection point. Applying the above error metric, the subdivision density is simply $\sqrt{|a(s-s_0)|}$. The integral is $\frac{2}{3}\sqrt{a}(s-s_0)^{1.5}$, which is readily invertible.

\begin{figure}
    \includegraphics[scale=0.3]{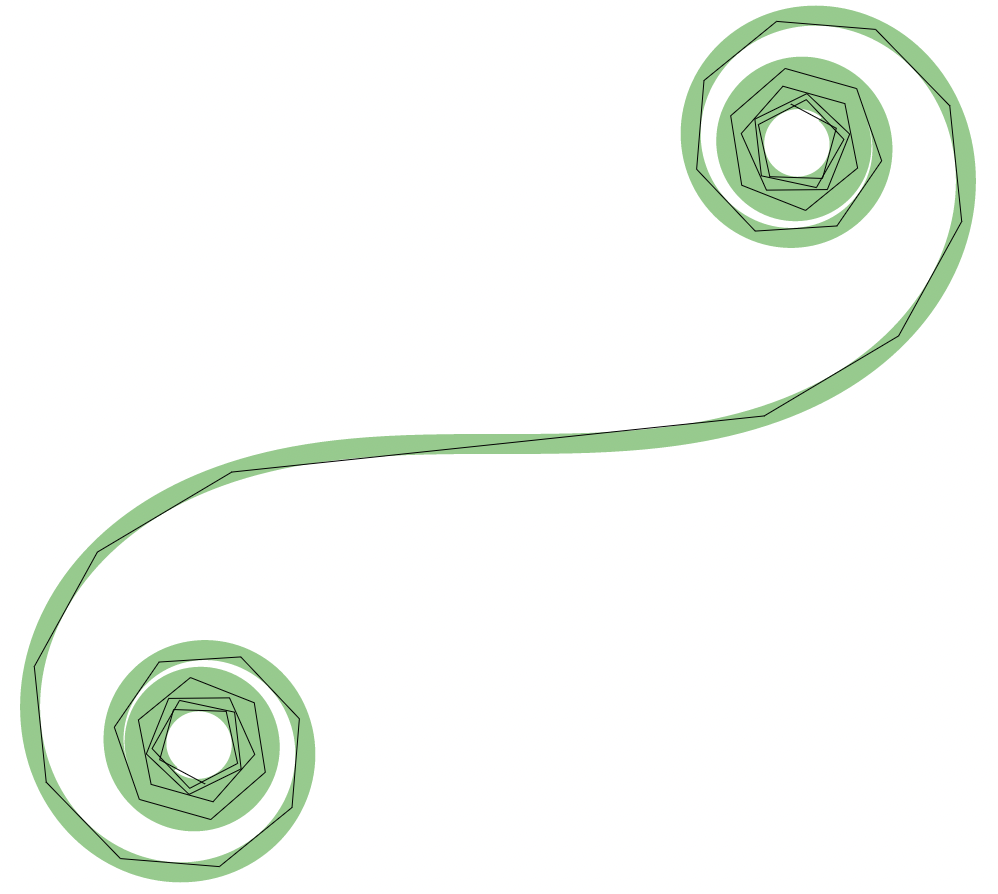}
    \caption{The flattened approximation to an Euler spiral using line segments is achieved by choosing subdivision points using a simple power law formula with an exponent of 2/3.}
    \Description{An Euler spiral along with its flattened representation. The flattened curve has a consistent error, and the subdivision points are spaced according to a power law.}
    \label{fig:es_flatten}
\end{figure}

An immediate consequence is that flattening an Euler spiral by choosing subdivision points $s_i = a\cdot i^\frac{2}{3}$ produces a near-optimal flattening, as visualized in Figure~\ref{fig:es_flatten}.

\section{Euler spirals and their parallel curves}

It is common to approximate cubic Béziers to some intermediate curve format more conducive to offsetting and flattening. A number of published solutions (\citet{Yzerman2020}, \citet{Nehab2020}) use quadratic Béziers, as it is well suited for computation of parallel curves. Even so, this curve has some disadvantages. For one, it cannot model an inflection point, so the source curve must be subdivided at inflection points.

Like these other approaches, we also use an intermediate curve, but our choice is an Euler spiral. In some ways it is similar to quadratic Béziers -- it also has $O(n^4)$ scaling and is best computed using geometric Hermite interpolation -- but differs in others. It has no difficulty modeling an inflection point. Further, its parallel curve has a particularly simple mathematical definition and clean behavior regarding cusps.

An Euler spiral segment is defined as having curvature linear in arc length.

The parallel curve of the Euler spiral (also known as ``clothoid'') was characterized by \citet{Wieleitner1907} well over a hundred years ago, and has a straightforward closed-form Cesàro representation, curvature as a function of arc length, of the form $\kappa(s) = c_0(s - s_0)^{-1/2} + c_1$. Euler spirals as an intermediate curve representation for computing parallel curves was proposed in \citet{Levien2021}, which also expands on the Wieleitner reference (including an English translation) and Cesàro equation.

\section{Flattened parallel curves}

The geometry of a stroke outline consists of joins, caps, and the two parallel curves on either side of the input path segments, offset by the half linewidth. The joins and caps are not particularly difficult to calculate, but parallel curves of cubic Béziers are notoriously tricky~[\citenum{Kilgard2020a}]. Analytically, it is a tenth order algebraic curve, which is not particularly feasible to compute directly.

Conceptually, generating a flattened stroke outline consists of computing the parallel curve of the input curve segment followed by \emph{flattening,} the generation of a polyline that approximates the parallel curve with sufficient accuracy (which can be measured as Fréchet distance). However, these two stages can be fused for additional performance, obviating the need to store a representation of the intermediate curve.

Using a subpixel Fréchet distance bound guarantees that the rendered image does not deviate visibly from the exact rendering. Another choice would be uniform steps in tangent angle, as chosen by polar stroking~[\citenum{Kilgard2020}]. However, at small curvature, the stroked path can be off by several pixels, and at large curvature there may be considerably more subdivision than needed for faithful rendering.

\begin{figure}
    \includegraphics[scale=0.5]{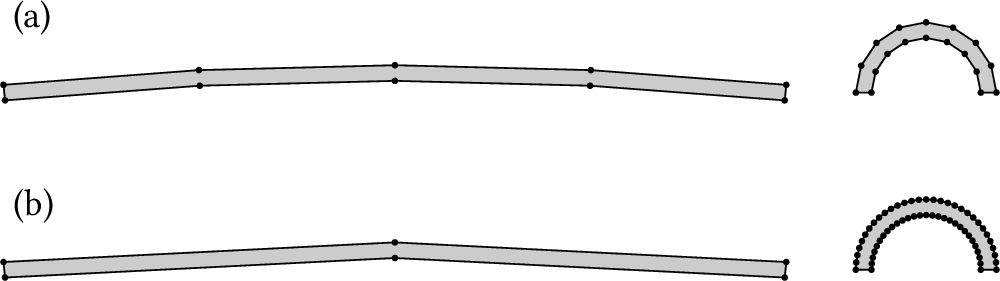}
    \caption{Comparison of error metrics. Top row (a) shows a distance based error metric. Bottom row (b) shows the angle step error metric.}
    \Description{Four stroked arcs with different amount of subdivision. The top row shows a low-curvature and high-curvature arc with a consistent distance error}
    \label{fig:angle_err}
\end{figure}

The limitation of the angle step error metric is shown in Figure~\ref{fig:angle_err}. The top row shows the use of a distance-based error metric, as is used in our approach, which is visually consistent at varying curvature (in practice, a tolerance value of 0.25 of a pixel is below the threshold of perceptibility). The bottom row shows a consistent angle step, as implemented in polar stroking, but has excessive distance error at low curvature, and excessive subdivision at high curvature. It should be noted, to avoid the undershoot at low curvature, both the Skia~[\citenum{Skia}] and Rive~[\citenum{Rive}] renderers use a hybrid of the Wang and polar stroking error metrics.

%
%

\subsection{The subdivision density integral} \label{subsection:subdiv-density-int}

\begin{figure}
    \includegraphics[scale=0.6]{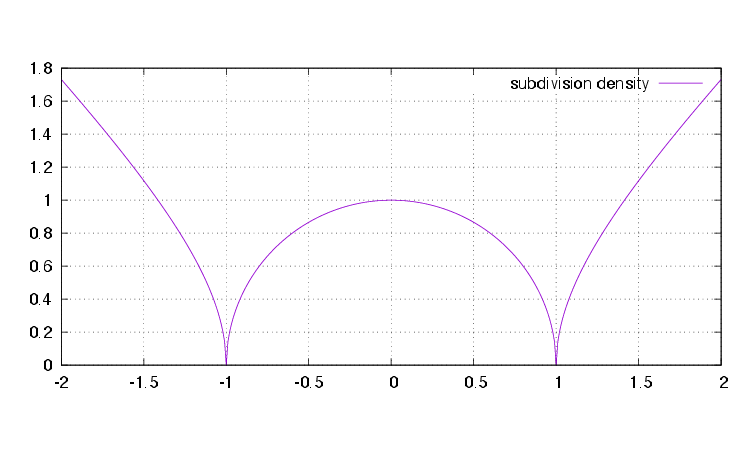}
    \caption{Subdivision density for the parallel curve of an Euler spiral}
    \Description{A graph of subdivision density. The range from -1 to 1 is a semicircle, and the value from 1 rises sharply from zero to an asymptote of unity slope.}
    \label{fig:subdiv_density}
\end{figure}

The subdivision density for the parallel curve of an Euler spiral, normalized so that its inflection point is at $-1$ and the cusp of the parallel curve is at 1, is simply $\sqrt{|1-s^2|}$. This function is plotted in Figure~\ref{fig:subdiv_density}.

The subdivision density integral for the parallel curve of an Euler spiral is given as follows:

\[
    f(x) = \int_0^x\sqrt{|u^2 - 1|} du
\]

This integral has a closed-form analytic solution:

\[
    f(x) = \left\{
        \begin{array}{rl}
            \frac{1}{2}(x\sqrt{|x^2-1|} + \sin^{-1}x) & \text{if } |x| \leq 1 \\
            \frac{1}{2}(x\sqrt{|x^2-1|} - \cosh^{-1}x + \frac{\pi}{4}) & \text{if } x \geq 1
        \end{array}
    \right.
\]

Values for $x<-1$ follow from the odd symmetry of the function.

\subsection{Approximation of the subdivision density integral}

\begin{figure}
    \includegraphics[scale=0.8]{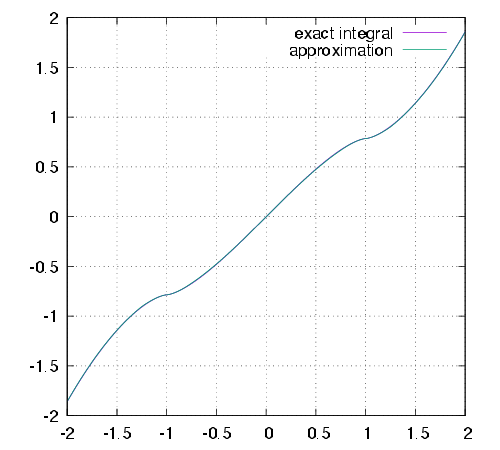}
    \caption{Integral of subdivision density for Euler spiral parallel curve, and its approximation}
    \Description{A graph of the integral of subdivision density. It is a wriggly diagonal line, with flat parts at -1 and 1. The approximation is nearly indistinguishable from the exact value.}
    \label{fig:espc}
\end{figure}

The subdivision density integral (Section~\ref{subsection:subdiv-density-int}) is fairly straightforward to compute in the forward direction, but not invertible using a straightforward closed-form equation. Numerical techniques are possible, but require multiple iterations to achieve sufficient accuracy, so are slower. In this subsection, we present a straightforward and accurate approximation, constructed piecewise from easily invertible functions. This approximation was found empirically, by interactively tuning candidate approximation functions in the Desmos graphing calculator. The exact integral and the approximation are shown in Figure~\ref{fig:espc}. Visually, it is clear that the agreement is close. We also built a test suite (included in our supplemental materials) to exhaustively test the subdivision count using a property testing approach, finding that the worst case discrepancy between approximate and exact results is 6\%. If higher flattening quality is desired at the expense of slower computation, this approximation could be used to determine a good initial value for numeric techniques; two iterations of Newton solving are enough to refine this guess to within 32-bit floating point accuracy.

The approximation is as follows:

\[ 
    f_\mathit{approx}(x) = \left\{
        \begin{array}{rl}
            \frac{\sin c_1 x}{c_1} & \text{if } x < 0.8 \\
            \frac{\sqrt{8}}{3}(x-1)^{1.5} + \frac{\pi}{4} & \text{if } 0.8 \leq x < 1.25 \\
            0.6406x^2 - 0.81x + c_2 & \text{if } 1.25 \leq x < 2.1 \\
            0.5x^2 - 0.156x + c_3 & \text{if } x \geq 2.1
        \end{array}
        \right.
\]
\[
    \begin{array}{ll}
        c_1 = & 1.0976991822760038 \\
        c_2 = & 0.9148117935952064 \\
        c_3 = & 0.16145779359520596
    \end{array}
\]

The primary rationale for the constants is for the approximation to be continuous. The other parameters were determined empirically; further automated optimization is possible but is unlikely to result in dramatic improvement. Further, this approximation is given for positive values. Negative values follow by symmetry, as the function is odd.

\section{Error metrics for approximation by arcs}

The problem of approximating a curve by a sequence of arc segments has extensive literature, but none of the published solutions are quite suitable for our application. The specific problem of approximating an Euler spiral by arcs is considered in \citet{Meek2004} using a ``cut then measure'' adaptive subdivision scheme, but their solution has poor quality; it scales as $O(1/n^2)$, while $O(1/n^3)$ is attainable. The result was improved ``slightly'' by \citet{Narayan2014}. The literature also contains optimal results, namely \citet{Maier2014} and \citet{Nuntawisuttiwong2021}, but at considerable cost; both approaches claim $O(n^2)$ time complexity. The through-line for all these results is that they are solving a harder problem: adopting the constraint that the generated arc sequence is $G^1$ continuous. While desirable for many applications, this constraint is not needed for rendering a stroke outline. Even with this constraint relaxed, the angle discontinuities of an arc approximation are tiny compared to flattening to lines.


Our approach is based on a simple error metric, similar in flavor to the one for flattening to line segments. The details of the metric (in particular, tuning of constants) were obtained empirically, though we suspect that more rigorous analytic bounds could be obtained. In practice it works very well indeed; the best way to observe that is an interactive testing tool, which is provided in the supplemental materials.

The proposed error metric is as follows. The estimated distance error for a curve of length $\hat{s}$ is:

\[
    d \approx \frac{1}{120}\left(\int_0^{\hat{s}} \sqrt[3]{|\kappa'(s)|}ds \right)^3
\]

For an Euler spiral segment, $\kappa'(s)$ is constant and thus this error metric becomes nearly trivial. With $n$ subdivisions, the estimated distance is simply $\frac{s^3\kappa'}{120n^3}$. Solving for $n$, we get $n = s\sqrt[3]{\frac{|\kappa'|}{120d}}$ subdivisions, and those are divided evenly by arc length, as the subdivision density is constant across the curve, just as is the case for flattening arcs to lines.

Remarkably, the approximation of an Euler spiral parallel curve by arc segments is almost as simple as that for Euler spirals to arcs. As in flattening to lines, the parameter for the curve is the arc length of the originating Euler spiral. The subdivision density is then constant, and only a small tweak is needed to the formula for computing the number of subdivisions, taking into account the additional curvature variation from the offset by $h$ (the half line-width). The revised formula is:

\[
    n = s\sqrt[3]{\frac{|\kappa'|(1+0.4|hs\kappa'|)}{120d}}
\]

This formula was determined empirically by curve-fitting measured error values from approximating Euler spiral parallel curves to arcs, but was also inspired by applying the general error metric formula to the analytical equations for Euler spiral parallel curve, and dropping higher order terms. A more rigorous derivation, ideally with firm error bounds, remains as future work.

One consequence of this formula is that, since the error is in terms of the absolute value of $h$, independent of sign, the same arc approximation can be used for both sides of a stroke.

\begin{figure}
    \includegraphics[scale=0.65]{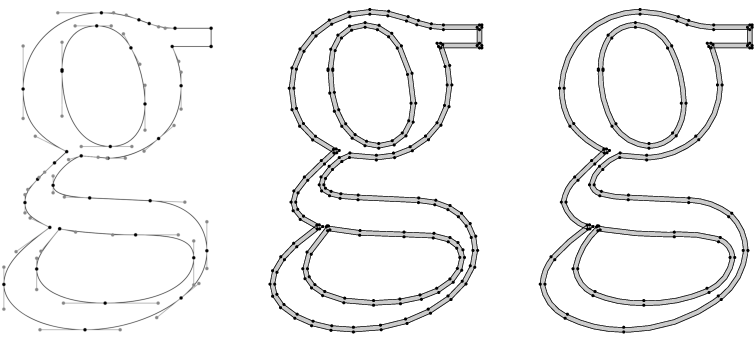}
    \caption{A lowercase `g' glyph from Nimbus Roman (left, constructed from cubic Bézier segments), its flattening to lines (center), and its approximation by arc segments (right), both with a tolerance of 2.0.}
    \Description{TODO}
    \label{fig:g_comparison}
\end{figure}

See Figure~\ref{fig:g_comparison} for a comparison between flattening to a polyline and approximation with arc segments. The arc segment version has many fewer segments at the same tolerance, while preserving very high visual quality.

\section{Evolutes}

In the principled, correct specification for stroking~[\citenum{Nehab2020}], parallel curves are sufficient only for segments in which the curvature does not exceed the reciprocal half-width. When it does, additional segments must be drawn, including evolutes of the original curve. In general, the evolute of a cubic Bézier is a very complex curve, requiring approximation techniques. By contrast, the evolute of an Euler spiral ($\kappa = as$) is another spiral with a simple Cesàro equation, namely $\kappa = -a^{-1}s^{-3}$, an instance of the general result that the evolute of a log-aesthetic curve is another log-aesthetic curve~[\citenum{Yoshida2012}].

Flattening this evolute is also straightforward; the subdivision density is proportional to $s^{-0.5}$ where $s$ is the arc length parameter of the underlying Euler spiral (and translated so $s = 0$ is the inflection point). Thus, the integral is $2\sqrt{s}$, and the inverse integral is just squaring. Thus, flattening the evolute of an Euler spiral is simpler than flattening its parallel curve.

\begin{figure}
    \includegraphics[scale=0.6]{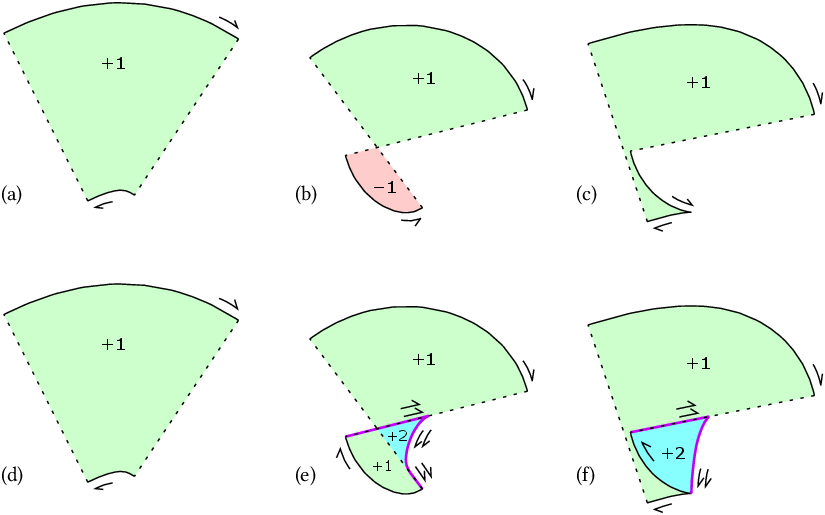}
    \caption{Weakly and strongly correct stroke outlines. The top row shows weakly correct stroke outlines. In (a) the curvature does not exceed the reciprocal half-width, and the stroke is rendered correctly. In (b) the curvature of the bottom outline consistently exceeds the reciprocal half-width, and a section of the outline incorrectly has negative winding number. In (c) there is a cusp. The bottom row shows corresponding strongly correct renders; (d) is the same as (a), while the other two show additional segments for the evolute and connecting lines (in purple). Note that the sections of parallel curve outline with above-threshold curvature are reversed, and that the section enclosed by the evolute has a total winding number of +2.}
    \Description{Two rows of rendered strokes, the top being weakly correct, only drawing parallel curves, the bottom being strongly correct, with additional evolute segments.}
    \label{fig:evolutes}
\end{figure}

The effect of adding evolutes to achieve strong correctness is shown in Figure~\ref{fig:evolutes}. The additional evolute segments and connecting lines are output twice, to make the winding numbers consistent and produce a watertight outline. All winding numbers are positive, so rendering with the nonzero winding rule yields a correct final render.

\section{Conversion from cubic Béziers to Euler spirals}

The Euler spiral segment representation of a curve is useful for computing near-optimal flattened parallel curves, but standard APIs and document formats overwhelmingly prefer cubic Béziers as the path representation.

Many techniques for stroke expansion described in the literature apply some lowering of cubic Bézier curves to a simpler curve type that is more tractable for evaluating the parallel curve. Computing parallel curves directly on cubic Bézier curve segments is not very tractable. In particular, the widely cited Tiller-Hanson algorithm~[\citenum{Tiller1984}] performs well for quadratic Béziers but significantly worse for cubics.


A typical pattern for converting from one curve type to another is \emph{adaptive subdivision.} An approximate curve is found in the parameter space of the target curve family. The error of the approximation is measured. If the error exceeds the specified tolerance, the curve is subdivided (typically at $t = 0.5$), otherwise the approximation is accepted. Subdivisions are also indicated at special points; for example, since quadratic Béziers cannot represent inflection points, and geometric Hermite interpolation is numerically unstable if the input curve is not convex, lowering to quadratic Béziers also requires calculation of inflection points and subdividing there. A good example of this pattern is \citet{Nehab2020}. One advantage of Euler spirals over quadratic Béziers is that they can represent inflection points just fine, so it is not necessary to solve for the inflection points, or do additional subdivision. Avoiding these alleviates the need for conditional logic and case analysis, which makes the algorithm more amenable to GPU execution.

The approach in this paper is another variant of adaptive subdivision, with two twists. First, it's not necessary to actually generate the approximate curve to measure the error. Rather, a straightforward closed-form formula accurately predicts it. The second twist is that, since compute shader languages on GPUs typically don't support recursion, the stack is represented explicitly and the conceptual recursion is represented as iterative control flow. This is an entirely standard technique, but with a clever encoding the entire state of the stack can be represented in two words, each level of the stack requiring a mere single bit.

\subsection{Error prediction}

A key step in approximating one curve with another is evaluating the error of the approximation. A common approach (used in \citet{Nehab2020} among others) is to generate the approximate curve, then measure the distance, often by sampling at multiple points on the curve. All this is potentially slow, with the additional risk of underestimating the error due to undersampling.

Our approach is different. In short, we perform a straightforward analytical computation to accurately estimate the error. Our approach to the error metric has two major facets. First, we obtain a secondary cubic Bézier which is a very good fit to the Euler spiral, then we estimate the distance between that and the source cubic. Due to the triangle inequality, the sum of these is a conservative estimate of the true Fréchet distance between the cubic and the Euler spiral.

For mathematical convenience, the error estimation is done with the chord normalized to unit distance; the actual error is scaled linearly by the actual chord length.

The cubic Bézier approximating the Euler spiral is one in which the distance of each control point from the endpoint is $d = \frac{2}{3(1 + \cos \theta)}$, where $\theta$ is the angle of the endpoint tangent relative to the chord. This is a generalization of the standard approximation of an arc. It should be noted that this is not in general the closest possible fit, but it is computationally tractable and has near-uniform parametrization. In general it has quintic scaling; if an Euler spiral segment is divided in half, the error of this cubic fit decreases by a factor of 32. A good estimate of this error is:

\[
    4.6255\times10^{-6}|\theta_0+\theta_1|^5 + 7.5\times10^{-3}|\theta_0+\theta_1|^2|\theta_0-\theta_1|
\]

The distance between two cubic Bézier segments can be further broken down into two terms; in most cases, the difference in area accurately predicts the Fréchet distance between the curves. Two cubic Béziers with the same area and same tangents tend to be relatively close to each other, but there is some error stemming from the difference in parametrization. Area of a cubic Bézier segment is straightforward to compute using Green's theorem:

\[
    a = \tfrac{3}{20}(2d_0\sin \theta_0 + 2d_1\sin \theta_1 - d_0 d_1\sin(\theta_0+\theta_1))
\]

The conservative Fréchet distance estimate is 1.55 the absolute difference in area between the source cubic and the Euler spiral approximation. The final term for imbalance is as follows, with $\bar{d}_0$ and $\bar{d}_1$ representing the distance from the endpoints of the Euler approximation and $d_0$ and $d_1$ the corresponding distance in the source cubic segment, as in the area calculation above:

\[
    (0.005|\theta_0+\theta_1| + 0.07|\theta_0 - \theta_1|)\sqrt{(\bar{d}_0 - d_0)^2 + (\bar{d}_1 - d_1)^2}
\]

The total estimated error is the sum of these three terms. We have validated this error metric in randomized testing. For values of $\theta$ between $0$ and $0.5$, and values of $d$ between $0$ and $0.6$, this estimate is always conservative, and it is also tight: over Bézier curves generated randomly with parameters drawn from a uniform distribution in this range, the mean ratio of estimated to true error is 1.656.

\begin{figure}
    \includegraphics[scale=0.75]{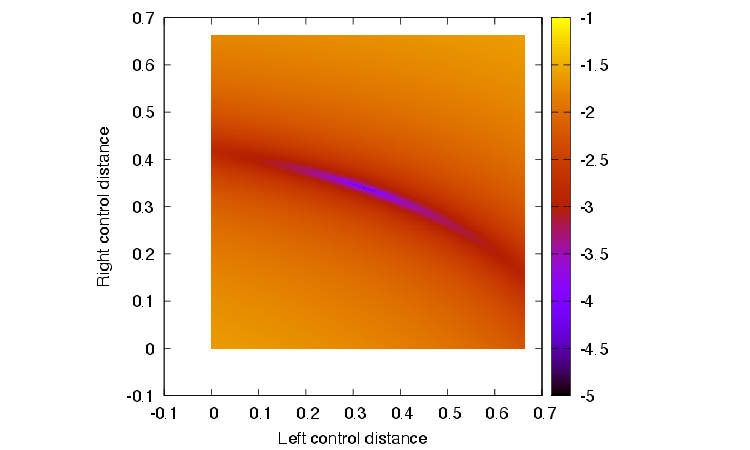} \includegraphics[scale=0.75]{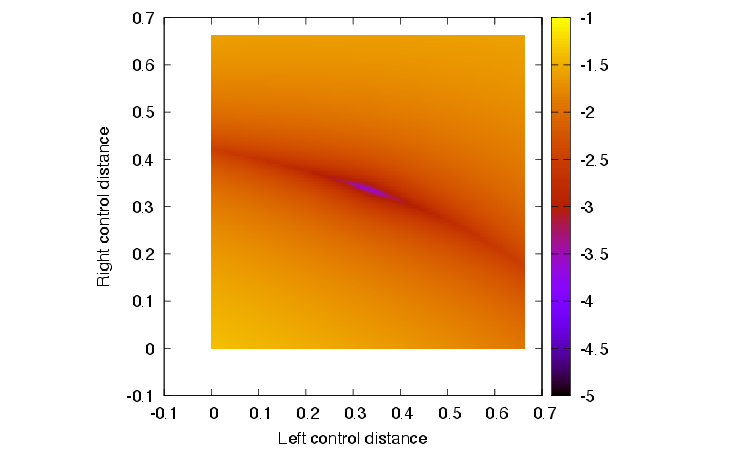}
    \caption{Comparison of measured (top) and approximated (bottom) error for cubic Bézier to Euler spiral conversion. Scale is log (base 10) of the error.}
    \Description{TODO}
    \label{fig:cubic_err}
\end{figure}

A visualization of combined error metric is shown in Figure~\ref{fig:cubic_err}, comparing measured and approximate error for a slice of the parameter space, fixing the endpoint angles to 0.1 and 0.2, and varying the distance from both endpoints to the control point.

\subsection{Geometric Hermite interpolation}

Given tangent angles relative to the chord, finding the Euler spiral segment that minimizes total curvature variation is a form of geometric Hermite interpolation. There are a number of published solutions to this problem, involving nontrivial numerical solving techniques: gradient descent~[\citenum{Kimia2003}], bisection~[\citenum{Walton2009}], or Newton iteration~[\citenum{Connor2014}]. A more direct approach is to approximate the function as a reasonably low-order polynomial in terms of the endpoint angles. Our approach is to use a 7th order polynomial, which is more precise than the 3rd order polynomial proposed in \citet{Reif2021}, at only modest increased cost. The exact coefficients used are presented in Appendix \ref{appendix:gh}.

\subsection{Cusp handling}

There are two types of cusps that must be handled in the stroke expansion problem. One is when the input cubic Bézier contains a cusp (or near-cusp, which is prone to numerical robustness issues), and one is when the parallel curve contains cusps. This section will describe both in order.

A Bézier curve is expressed in parametric form, and its derivative with respect to the parameter can be zero or nearly so, causing serious numerical problems for many stroke expansion algorithms. In the limit, as the Bézier curve describes a semi-cubical parabola, the curvature can become infinite.

Our approach leverages the conversion to Euler spiral segments, which have finite curvature. The geometric Hermite interpolation depends on accurate tangents at the endpoints, which in turn are derived from derivatives. The tangents are not well defined when the derivative is zero, and are not numerically stable when near-zero. Thus, the algorithm has one major mechanism to deal with numerical robustness in these cases: when sampling the cubic Bézier, if the derivative is near zero (as determined by testing against an epsilon threshold), then the derivative is sampled at a slightly perturbed parameter value.

\begin{figure}
    \includegraphics[scale=0.24]{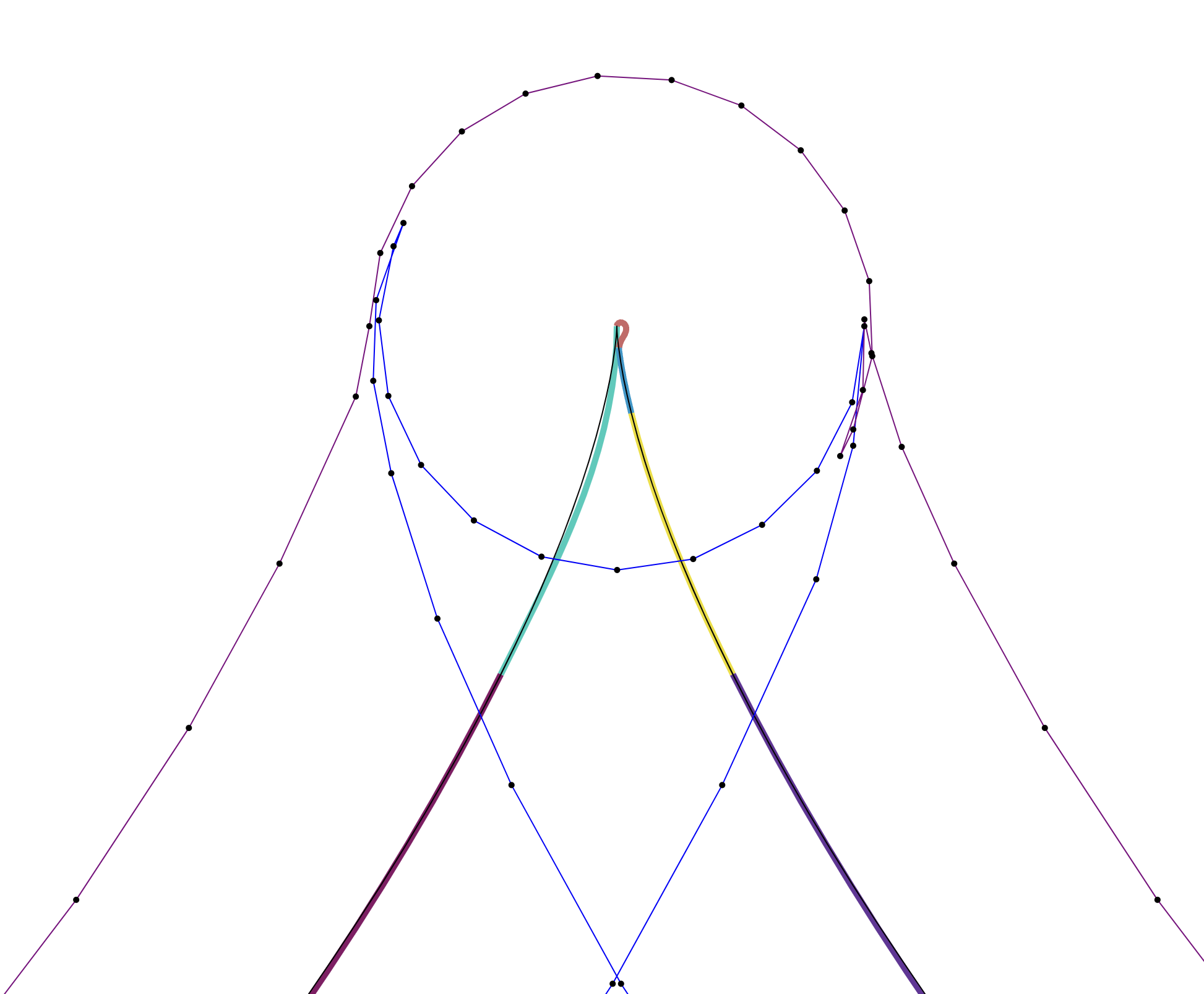}
    \caption{Rendering of a cubic Bézier cusp using Euler spiral segments}
    \Description{A cubic Bézier with a cusp is approximated by Euler spiral segments with finite curvature, including one with a question mark shape. The resulting stroke outlines have the correct shape.}
    \label{fig:cusp_rendering}
\end{figure}

In practice, when rendering a cubic Bézier with a cusp, the region near the cusp is rendered with one Euler spiral segment approximately in shape of the top of a question mark, as shown in Figure~\ref{fig:cusp_rendering}. Its parallel curve is well defined, and has the correct shape for the outline of the stroke, given of course that the distance is within tolerance (as enforced by the error metric).

\begin{figure*}
    \includegraphics[scale=0.19]{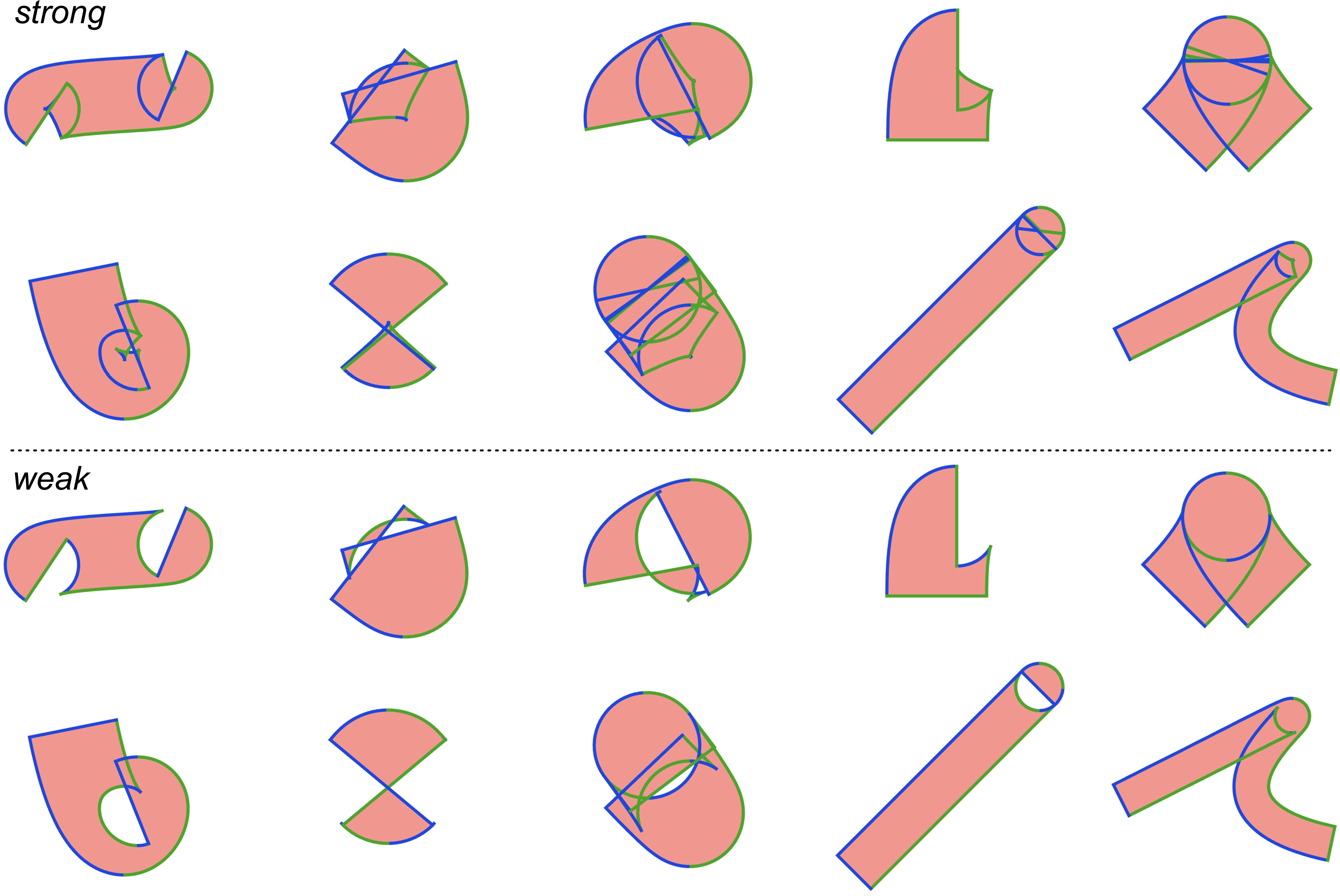}
    \caption{Stroked cubic Béziers exhibiting strong and weak correctness, rendered by our implementation. The contents of the line soup buffer are visualized as green and blue outlines. Segments with a winding direction that points \emph{up} are colored blue while those pointing \emph{down} are colored green. The curves were adapted from the \emph{stills} dataset by \citet{Nehab2020} and the Skia test corpus.}
    \Description{Selection of stroked cubic Béziers exhibiting strong and weak correctness}
    \label{fig:stroke_stills}
\end{figure*}

As recommended in both \citet{Nehab2020} and \citet{Kilgard2020}, the rendering of a cusp with infinite curvature matches that of a near-cusp. The code to detect near-zero derivatives and re-evaluate is a small amount of non-divergent logic, unlike the additional ``regularization'' pass proposed in Section 3.1 of \citet{Nehab2020} to handle special cases.

\subsubsection{Cusps in parallel curves}

Even when the input curve is smooth, its parallel curve contains a cusp when the (signed) curvature equals the offset distance (the half-width of the stroke). In published techniques, dealing with these cusps is a nontrivial effort, and involves numerical methods that are not GPU friendly. In particular, detecting locations in the cubic Bézier where the curvature crosses a given quantity is a medium-degree polynomial, and in general requires numerical techniques for root finding. In the approach of \citet{Nehab2020}, the root finder not only requires a hybrid Newton/bisection method, which requires iteration, but is also recursive in that it uses the roots of a polynomial of one degree lower as a subroutine. In general, polynomials up to cubic can be considered GPU-friendly, while finding cusps in cubic Béziers requires a higher degree.

In the simplest case, weak stroke correctness with line segments as the output primitive, no additional work is needed -- the flattening algorithm for Euler spiral parallel curves will naturally generate a point within the given error tolerance of the true cusp. However, generation of arc segments and drawing the evolutes needed for strong correctness both require subdivision at the parallel curve cusps.

Fortunately, finding the cusp in the parallel curve of an Euler spiral is a simple linear equation, and there is at most one cusp in any such segment. Euler spirals are thus a good solution for determining cusps as a piecewise linear approximation to curvature.

\section{GPU Implementation} \label{section:gpu-impl}

We applied the techniques outlined in Section 7 to implement a data-parallel algorithm that can convert stroked 2D Bézier paths into flat geometry suitable for GPU rendering. We focused primarily on the global stroke-to-fill conversion problem in order to generate polygonal outlines for rasterization. Since the Euler spiral approximation can fit any source cubic Bézier and its parallel curves, our implementation can be used to render both filled and stroked primitives within a satisfactory error tolerance.

We aimed to satisfy the following the criteria:
\begin{enumerate}
  \item The implementation must be able to handle a large number of inputs at real-time rates. 
  \item The stroke outlines must meet the strong correctness definition.
  \item The implementation must support the standard SVG stroke end cap (\emph{butt}, \emph{square}, \emph{round}) and join styles (\emph{bevel}, \emph{miter}, \emph{round}).
  \item The CPU must do no work beyond the basic preparation of the input path data for GPU consumption, in order to maximize the exploited parallelism.
\end{enumerate}

One of our artifacts is a GPU compute shader that satisfies these criteria. We coupled this with a simple and efficient data layout for parallel processing. Notably, our implementation is fully data-parallel on input path segments and does not require any computationally expensive curve evaluation on the CPU. We implemented our shader on general-purpose GPU compute primitives supported by all modern graphics APIs, particularly Metal, Vulkan, D3D12, and WebGPU~[\citenum{WebGPU}]. As such, the ideas presented in this section are portable to various GPU platforms.

The rest of this section describes the design of our GPU pipeline and input encoding scheme.

\subsection{Pipeline Design} \label{subsec:pipeline-design}

An SVG path consists of a sequence of instructions (or ``verbs''). The \emph{lineto} and \emph{curveto} instructions denote an individual \emph{path segment} consisting of either a straight line or a cubic Bézier. The \emph{moveto} instruction is used to begin a new \emph{subpath} and the \emph{closepath} instruction can be used to create a closed contour (by inserting a line connecting the current endpoint to the start of the subpath). A subpath must form a closed contour if painted as a fill. When painted as a stroke, each subpath must begin and end with a cap according to the desired cap style. In addition, each path segment in a stroked subpath must be connected to adjacent path segments of the same subpath with a join. For any given path, the cap and join styles do not vary across subpaths or individual path segments.

The standard organizational primitive for a compute shader is the \emph{workgroup}. A workgroup can be organized as a 1, 2, or 3-dimensional grid of individual threads that can cooperate using shared memory. Multiple workgroups can be \emph{dispatched} as part of a larger global grid, also of 1, 2, or 3 dimensions.

Our compute shader is arranged as a 1-dimensional grid, parallelized over input path segments. We chose a workgroup size of 256 as it works well over a wide range of GPUs. However, this choice of workgroup size is not particularly important for our algorithm as it does not require any cross-thread communication. Each thread in the global grid is assigned to process a single path segment. We encode the individual path segments contiguously in a GPU storage buffer such that each element has 1:1 correspondence to a global thread ID in the dispatch. This layout easily scales to hundreds of thousands of input path segments, which can be distributed across any number of paths and subpaths. We submit the kernel to the GPU as a single dispatch with as many workgroups as needed to handle the entire input.

\begin{center}
\begin{figure}
    \includegraphics[scale=0.18]{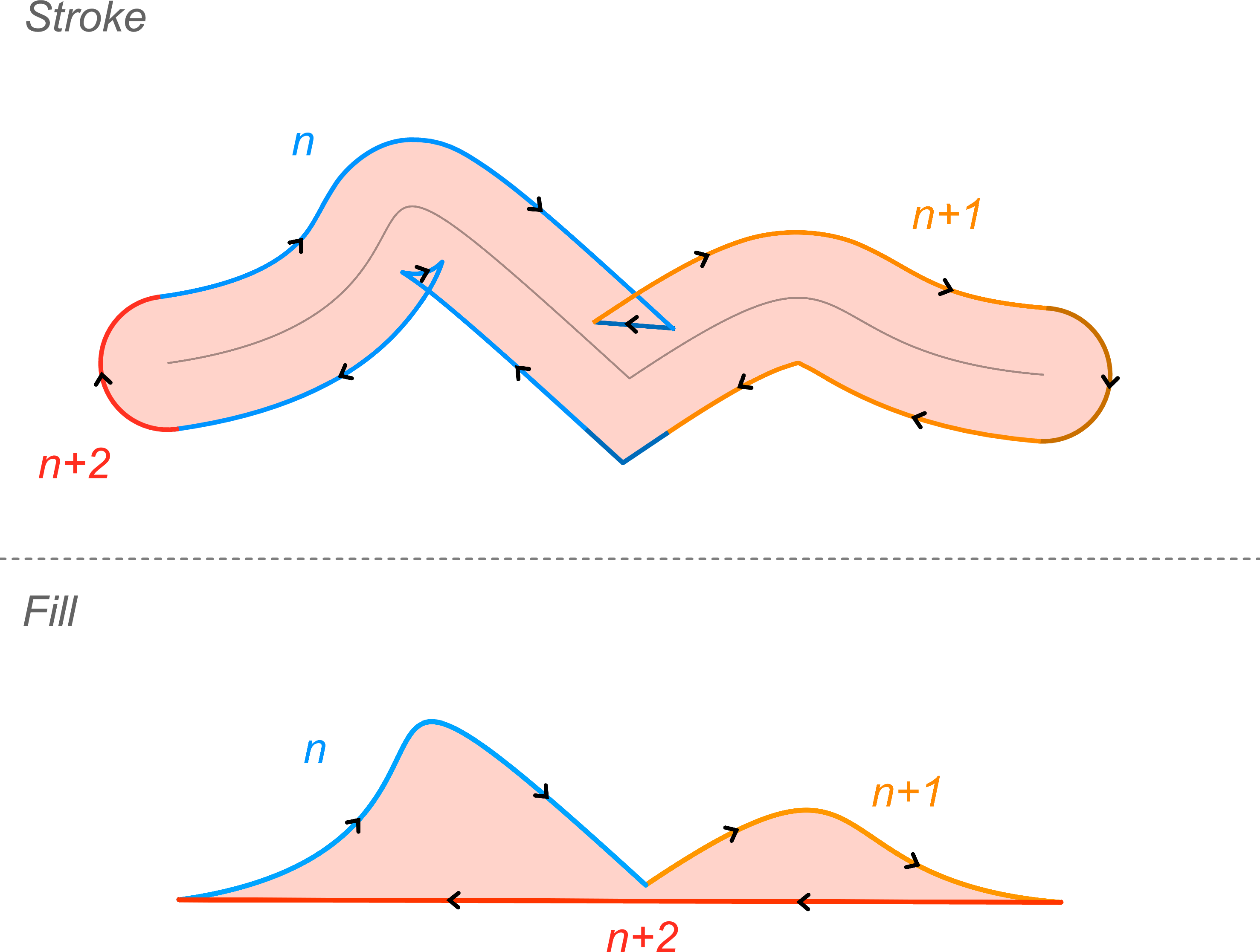}
    \caption{Visualization of contour outlines and the IDs of GPU threads that flatten them. The subpath contains two adjacent cubic Bézier curves rendered as both a stroke and a fill. In the stroked version, thread $n$ flattens the parallel curves of the first curve (blue) and the join segment with the miter style (dark blue). Thread $n + 1$ flattens the parallel curves of the second curve (orange) and the end cap with the round style (dark orange). Thread $n + 2$ handles the \emph{subpath end segment} and flattens the start cap (red). When the same path is rendered as a fill, threads $n$ and $n + 1$ each flatten the source curves, while thread $n + 2$ outputs the \emph{close} segment that joins the end of the subpath to its starting point. The black arrows represent the winding direction of the output segments.}
    \Description{TODO}
    \label{fig:stroke_threads}
\end{figure}
\end{center}

\begin{algorithm}
  $tid \gets$ global invocation ID\;
  $(style, segment) \gets readScene(tid)$\;
  \eIf{$style$ is $fill$}{
      \tcp{Output source curve (offset = 0)}
      flatten\_cubic(segment, $0$)\;
  }{
      \tcp{Handle stroked segment}
      \eIf{$segment$ is \emph{stroke cap marker}}{
          \eIf{path is open}{output start cap\;}{output nothing\;}
      }{
          \tcp{Output parallel curves}
          flatten\_cubic(segment,  style.offset)\;
          flatten\_cubic(segment, -style.offset)\;

          $neighbor \gets readScene(tid + 1)$\;
          \eIf{$neighbor$ is not \emph{stroke cap marker} OR path is closed}{
              output join\;
          }{
              output end cap\;
          }
      }
  }
\caption{The control flow of the compute shader}
\label{alg:shader-flow}
\end{algorithm}

Each thread outputs a polyline that fits a subset of the rendered subpath's outer contour, represented by the path segment (see Figure \ref{fig:stroke_threads}). For a filled primitive, a thread only outputs the flattened approximation of the path segment. For a stroke, a thread outputs the polylines approximating multiple curves: a) the two parallel curves on both sides of the source segment, offset by the desired stroke \emph{half-width}, b) the joins that connect the path segment to the next adjacent segment, and c) evolute patches in regions of high curvature. We also store metadata in our input encoding that marks the position of a path segment within the containing subpath. We use this information to determine whether to output an end cap instead of a join. We encode one additional segment designated as the \emph{stroke cap marker} for every subpath. The thread assigned to the marker segment only outputs a start cap, which completes the stroke outline.

The joins between adjacent segments need to form a continuous outline when combined with the segments. This requires that a thread have access to the tangent vector at the end of its assigned segment \emph{and} the start tangent of the next adjacent segment. We require that the input segments within a subpath are stored in order which simplifies the access to the adjacent segment down to a buffer read at the next array offset.

\newcommand{\nosemic}{\SetEndCharOfAlgoLine{\relax}}
\newcommand{\dosemic}{\SetEndCharOfAlgoLine{\string;}}
\newcommand{\pushline}{\Indp}
\newcommand{\popline}{\Indm\dosemic}

\begin{algorithm}
\SetKwProg{Fn}{fn}{}{end}
\SetKwBlock{Loop}{loop}{end}
\Fn{flatten\_cubic(cubic: Cubic, offset: f32)}{
    TOL $\gets$ Euler spiral fit tolerance\;
    \tcp{`t0\_u' and `dt' track the recursion stack}
    t0\_u $\gets$ 0\;
    dt $\gets$ 1.0\;
    last\_p $\gets$ cubic.p0\;
    last\_q $\gets$ cubic.p1 - cubic.p0\;
    \Loop{
        t0 $\gets$ f32(t0\_u) * dt\;
        \If{t0 == 1.0}{break\;}
        t1 $\gets$ t0 + dt\;
        (next\_p, next\_q) $\gets$ eval\_cubic\_and\_deriv(cubic, t1)\;
        \tcp{Estimate error of geometric}
        \tcp{Hermite interpolation to Euler spiral}
        \nosemic(th0, th1, chord\_len, err) $\gets$ fit\_cubic\_to\_es(\;
        \pushline\dosemic last\_p, next\_p, last\_q, next\_q, dt)\;
        \popline \eIf{err <= TOL}{
            es $\gets$ euler\_segment(th0, th1, last\_p, next\_p)\;
            \tcp{Find location of any cusp in outline.}
            \tcp{If $0 \leq t < 1$, draw evolute at `t'.}
            cusp0 $\gets$ es\_curvature(es, 0) * offset + chord\_len\;
            cusp1 $\gets$ es\_curvature(es, 1) * offset + chord\_len\;
            t $\gets$ cusp0 * cusp1 >= 0 ? 1 : cusp0 / (cusp0 - cusp1)\;
            \If{cusp0 >= 0}{
                evolute\_finalize()\;
                es\_seg\_flatten\_offset(es, offset, vec2(0, t))\;
            }
            \If{cusp0 < 0 OR t < 1}{
                evolute\_start(es, offset)\;
                range $\gets$ cusp0 >= 0 ? vec2(t, 1) : vec2(0, t)\;
                es\_seg\_flatten\_evolute(es, range)\;
                es\_seg\_flatten\_offset\_reverse(es, offset, range)\;
            }
            \If{cusp0 < 0 AND t < 1}{
                evolute\_finalize()\;
                es\_seg\_flatten\_offset(es, offset, vec2(t, 1))\;
            }
            \tcp{Pop the stack}
            t0\_u $\gets$ t0\_u + 1\;
            shift $\gets$ countTrailingZeros(t0\_u)\;
            t0\_u $\gets$ t0\_u $\gg$ shift\;
            dt $\gets$ dt * f32(1 $\ll$ shift)\;
            last\_p $\gets$ next\_p\;
            last\_q $\gets$ next\_q\;
        }{
            \tcp{Error is above the tolerance.}
            \tcp{Push the stack; Continue subdivision}
            t0\_u $\gets$ t0\_u * 2\;
            dt $\gets$ dt * 0.5\;
        }
    }
}
\caption{Outline of the iterative cubic Bézier flattening routine invoked in our compute shader. A working WGSL implementation is provided in the supplemental materials.}
\label{alg:flatten-cubic}
\end{algorithm}

All input segments first get converted to a cubic Bézier by degree elevation. Our flattening routine (see Algorithm~\ref{alg:flatten-cubic}) starts by computing the error estimate for geometric Hermite interpolation from the cubic Bézier to an Euler spiral segment. If the error exceeds the desired tolerance threshold, the cubic segment is iteratively subdivided in half and a new error estimate is computed. Once the error satisfies the tolerance, the Euler spiral segment is directly flattened to a polyline by invoking a subroutine that is parameterized by the stroke half-width (see \texttt{es\_seg\_flatten\_offset} in Algoritm~\ref{alg:flatten-cubic}). For a regular fill, this parameter is set to $0$. For strokes, the sign of the parameter determines the position of the parallel curve relative to the source segment as well as the winding direction of the output lines.

The adaptive subdivision in lowering cubic Béziers to Euler spiral segments is traditionally accomplished through recursion; in the case of subdivision there is a recursive call for each of the two subdivisions of the range. As the execution model of compute shaders does not support recursion, we \emph{unroll} recursion into an iterative loop. The cookbook technique for such unrolling is to store the stack explicitly as an array, which is potentially expensive (it increases pressure on registers). We exploit the specific nature of this recursion, encoding the entire state of the stack into two scalar values: the size of the range $dt$, and the scaled start of the range $t0\_u$, so that the range is $t0\_u \cdot dt .. (t0\_u + 1) \cdot dt$. Initial values are $1$ and $0$ respectively. Pushing the call stack is represented by halving $dt$ and doubling $t0\_u$, which preserves the start of the range but halves the size. Accepting an approximation (a leaf in the call stack) is represented by incrementing $t0\_u$. At this point, the decision of whether to pop the stack or continue is dependent on whether $t0\_u$ is even or odd, respectively. If even, popping the stack is accomplished by doubling $dt$ and halving $t0\_u$, and this is repeated until $t0\_u$ becomes odd. The repeated stack popping can be accomplished without iteration by using the ``count trailing zeros'' intrinsic to determine the number of stack pops $n$, then multiplying $dt$ by $2^n$, and shifting $t0\_u$ right by $n$ bits.

This procedure is sufficient to generate weakly correct stroke outlines. If the outline contains a cusp, we output an evolute patch (as shown in Figure~\ref{fig:evolutes}) in order to achieve strong correctness. A selection of renderings produced by our algorithm that demonstrate both strong and weak correctness is shown in Figure~\ref{fig:stroke_stills}.

The output of the compute shader is a collection of line segments that gets stored in a GPU storage buffer. Every element in the output contains the viewport coordinates of the two endpoints of a single line segment. Threads reserve storage buffer regions for their output by incrementing an atomic index stored in GPU device memory. A property of the data-parallel structure is that the line segments are unordered. Because the segments are processed in parallel, ordering cannot be guaranteed across path segments. We refer to this output data structure as \emph{line soup}.

\begin{figure*}
    \includegraphics[scale=0.275]{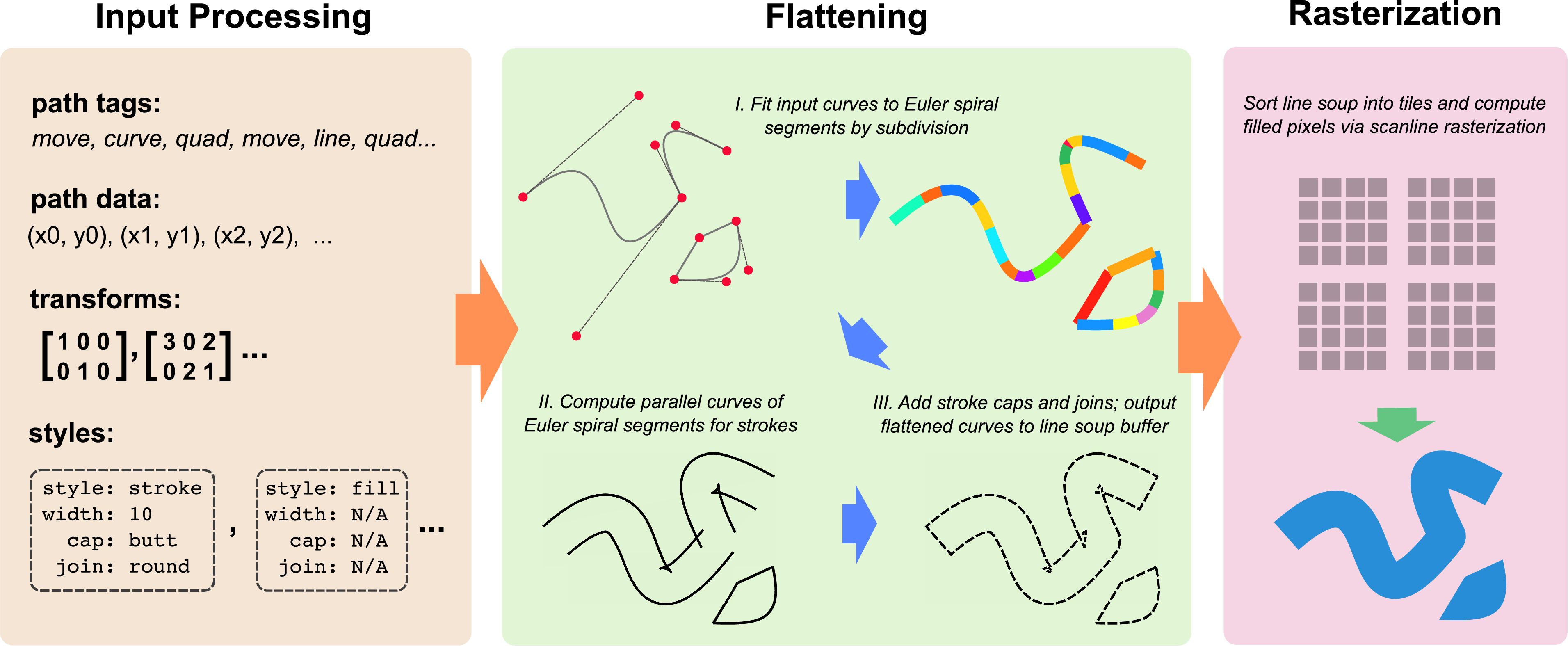}
    \caption{We integrated our curve flattening shader into our own renderer implemented entirely in compute shaders. The flattening stage (described in Section~\ref{section:gpu-impl}) converts stroked paths to filled outlines and flattens them to an unordered soup of line segments. Our renderer spatially sorts the line soup into tiles before computing pixel coverage using a highly parallel scanline rasterizer.}
    \Description{Our GPU pipeline architecture in broad strokes}
    \label{fig:pipeline}
\end{figure*}

An application may want to sort the line soup, depending on the requirements of the chosen rendering algorithm. For instance, our own renderer employs tile-based scanline rasterization in a compute shader, with a pipeline architecture that resembles \emph{cudaraster}~[\citenum{Laine2011}] and \emph{MPVG}~[\citenum{Ganacim2014}]. In this design, the line soup elements are spatially sorted by subsequent compute stages into a grid of 16x16 pixel tiles. The final compute stage of our renderer (called \emph{fine rasterizer}) operates at tile granularity and computes pixel coverage for filled polygons outlined by the line soup (Figure~\ref{fig:pipeline} illustrates the overall structure of our renderer, the details of which are beyond the scope of this paper).

We also implemented a variant that outputs circular arc segments instead of lines, in which the output entries store curvature in addition to the endpoints. For both primitive types, our renderer also outputs a 32-bit \emph{path ID} used by downstream pipeline stages to identify the path that a segment belongs to, in order to retrieve the correct paint (solid or gradient color) for rendering.

\subsection{Input encoding}

The input to our pipeline consists of a sequence of paths, each with an associated 2D affine transformation matrix and a \emph{style} data structure containing information about the stroke style or fill rule. We encode paths, transforms, and styles as parallel input streams. Unlike a conventional structure-of-arrays arrangement, we permit a one-to-many mapping across elements between certain streams: each transform and style entry can apply to one or more path entries with an array index greater than or equal to transform or style index.

Paths get encoded as two parallel streams: a \emph{path tag} stream containing an element for every segment in every subpath, and a \emph{path data} stream of point coordinates. Each segment/verb contributes a variable number of points: 1 for \emph{moveto}/\emph{lineto}, 2 for quadratic Béziers, and 3 for cubic Béziers. This variable length stream encoding allows for a highly compact representation of SVG-type content in memory. A GPU thread assigned to a path segment must be able to reference the corresponding element in each stream in order to obtain the right curve coordinates, transform, and style information. This is done by computing a set of stream offsets for every element in the path tag stream.


We compute the offsets on the GPU in a separate compute shader that runs prior to the flattening stage. A path tag consists of 8 bits that store \emph{stream offset increments}, such that the inclusive prefix sums of all increments for every element in the tag stream yield the stream offsets. We call this structure the \emph{tag monoid} and have implemented the computation as a parallel prefix scan.

\begin{table}
    \caption{
        \emph{Coordinate count} indicates the type of path segment: 1 for lines, 2 for quadratic Béziers, 3 for cubic Béziers. The \emph{subpath end bit} is set on the last segment of a subpath and the \emph{path bit} is additionally set on the last tag of a path. The number of curve control points is given by \emph{coordinate count}, plus $1$ if the subpath end bit is set. The number of bytes per coordinate pair is 4 for 16 bit or 8 for 32 bit coordinates.
    }
    \begin{tabular}{|c|l|}
    \hline
        \textbf{Path Tag Bits} & \multicolumn{1}{|c|}{\textbf{Monoid Fields}} \\
    \hline
        0-1 & coordinate count \\
        2 & subpath end bit \\
        3 & 16/32 bit coordinates \\
        4 & path bit \\
        5 & transform bit \\
        6 & style bit \\
    \hline
    \end{tabular}
    \label{table:pathtag}
\end{table}

Table~\ref{table:pathtag} shows the individual fields of the 8-bit path tag. The offset increments for the transform and style streams can be stored in a single bit that is set to $1$ only if a path tag marks the beginning of a new transform and/or style entry. We also encode an additional \emph{path bit} in the final segment of all paths, which allows us to access additional streams (such as fill color or gradient parameters) using a per-path offset in later rendering stages.

The handling of the subpath bit allows for overlap in coordinates, so that except at subpath boundaries, the first coordinate pair of each segment overlaps with the last coordinate pair of the previous one. During the CPU-side encoding, subpath boundaries are moved from the beginning of a subpath (moveto) to the end. The encoding process inserts an additional line segment into the tag and coordinate streams to close the subpath if the final point does not coincide with the start point. The \emph{subpath end bit} is set to $1$ for the final segment of every subpath. For strokes, the subpath end segment represents the \emph{stroke cap marker} defined in Section \ref{subsec:pipeline-design}. This special segment is assigned a coordinate count of $2$, and the corresponding two points in the path data stream encode the tangent vector of the subpath's initial segment, used to correctly draw the start cap or the connecting join in a closed contour.


\section{Results}


We present two versions of our stroking method: a sequential CPU stroker and the GPU compute shader outlined in Section \ref{section:gpu-impl}. Both versions can generate stroked outlines with line and arc primitives. We focused our evaluation on the number of output primitives and execution time, using the \emph{timings} dataset provided by \citet{Nehab2020}, in addition to our own curve-intensive scenes to stress the GPU implementation. We present our findings in the remainder of this section.

\subsection{CPU: Primitive Count and Timing}

\begin{figure}
    \includegraphics[scale=0.6]{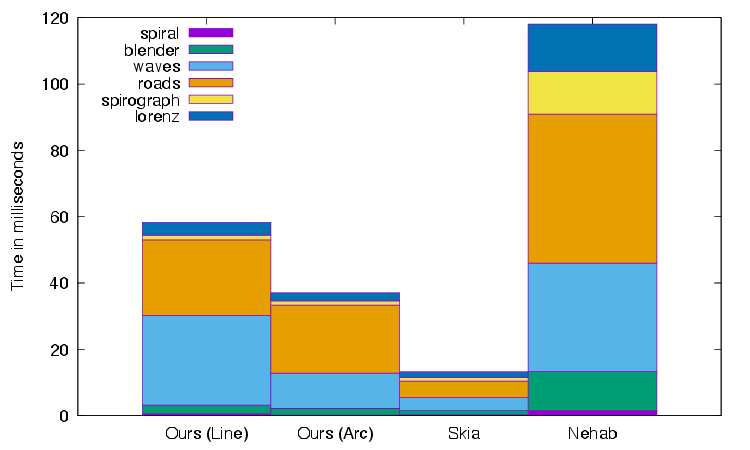}
    \caption{CPU timings for stroke expansion, comparing two versions of our stroker (outputting line and arc primitives) against Skia's curve-to-curve stroker. Measured on an AMD 3970}
    \Description{A stacked bar chart of the timings for stroke expansion}
    \label{fig:timings}
\end{figure}

We measured the timing of our CPU implementation, generating both line and arc primitives, against the Nehab and Skia strokers, which both output a mix of straight lines and quadratic Bézier primitives. The results are shown in Figure~\ref{fig:timings}. The time shown is the total for test files from the \emph{timings} dataset of \citet{Nehab2020}, and the tolerance was 0.25 when adjustable as a parameter.

Similarly to \citet{Nehab2020}'s observations, Skia is the fastest stroke expansion algorithm we measured. The speed is attained with some compromises, in particular an imprecise error estimation that can sometimes generate outlines exceeding the given tolerance. We confirmed that this behavior is still present. The Nehab paper proposes a more careful error metric, but a significant fraction of the total computation time is dedicated to evaluating it.

We also compared the number of primitives generated, for varying tolerance values from 1.0 to 0.001. The number of arcs generated is significantly less than the number of line primitives. The number of arcs and quadratic Bézier primitives is comparable at practical tolerances, but the count of quadratic Béziers becomes more favorable at finer tolerances. The vertical axis shows the sum of output segment counts for test files from the \emph{timings} dataset, plus mmark-70k (as described in more detail in Section~\ref{subsection:gpu-results}), and omitting the \emph{waves} example\footnote{The \emph{waves} example triggers Skia issue https://issues.skia.org/issues/336617138 at finer tolerance.}.

\begin{figure}
    \includegraphics[scale=0.6]{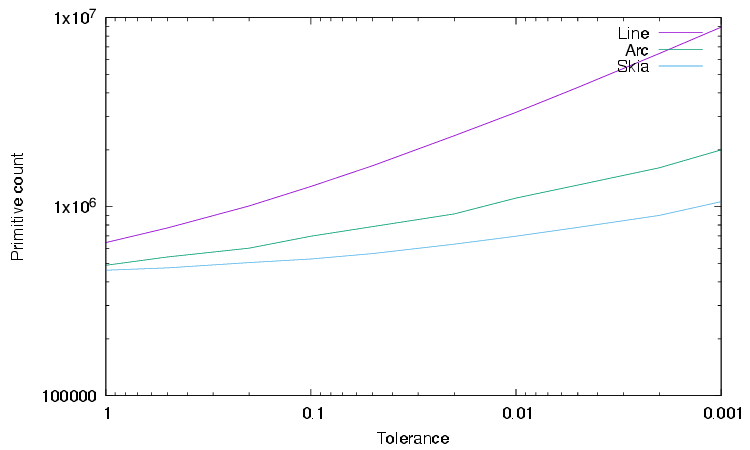}
    \caption{Primitive count for stroke expansion}
    \Description{A line chart showing the primitive count}
    \label{fig:prim_count}
\end{figure}

\subsection{GPU: Execution Time} \label{subsection:gpu-results}

We evaluated the runtime performance of our compute shader on 4 GPU models and multiple form factors: Google Pixel 6 equipped with the Arm \emph{Mali-G78 MP20} mobile GPU, the Apple \emph{M1 Max} integrated laptop GPU, and two discrete desktop GPUs: the mid-2015 NVIDIA {GeForce GTX 980Ti} and the late-2022 NVIDIA {GeForce RTX 4090}. We authored our entire GPU pipeline in the \emph{WebGPU Shading Language} (WGSL) and used the \emph{wgpu}~[\citenum{wgpu}] framework to interoperate with the native graphics APIs (we tested the M1 Max on macOS via the Metal API; the mobile and desktop GPUs were tested via the Vulkan API on Android and Ubuntu systems).

\begin{figure*}
    \centering
    \includegraphics[scale=0.2]{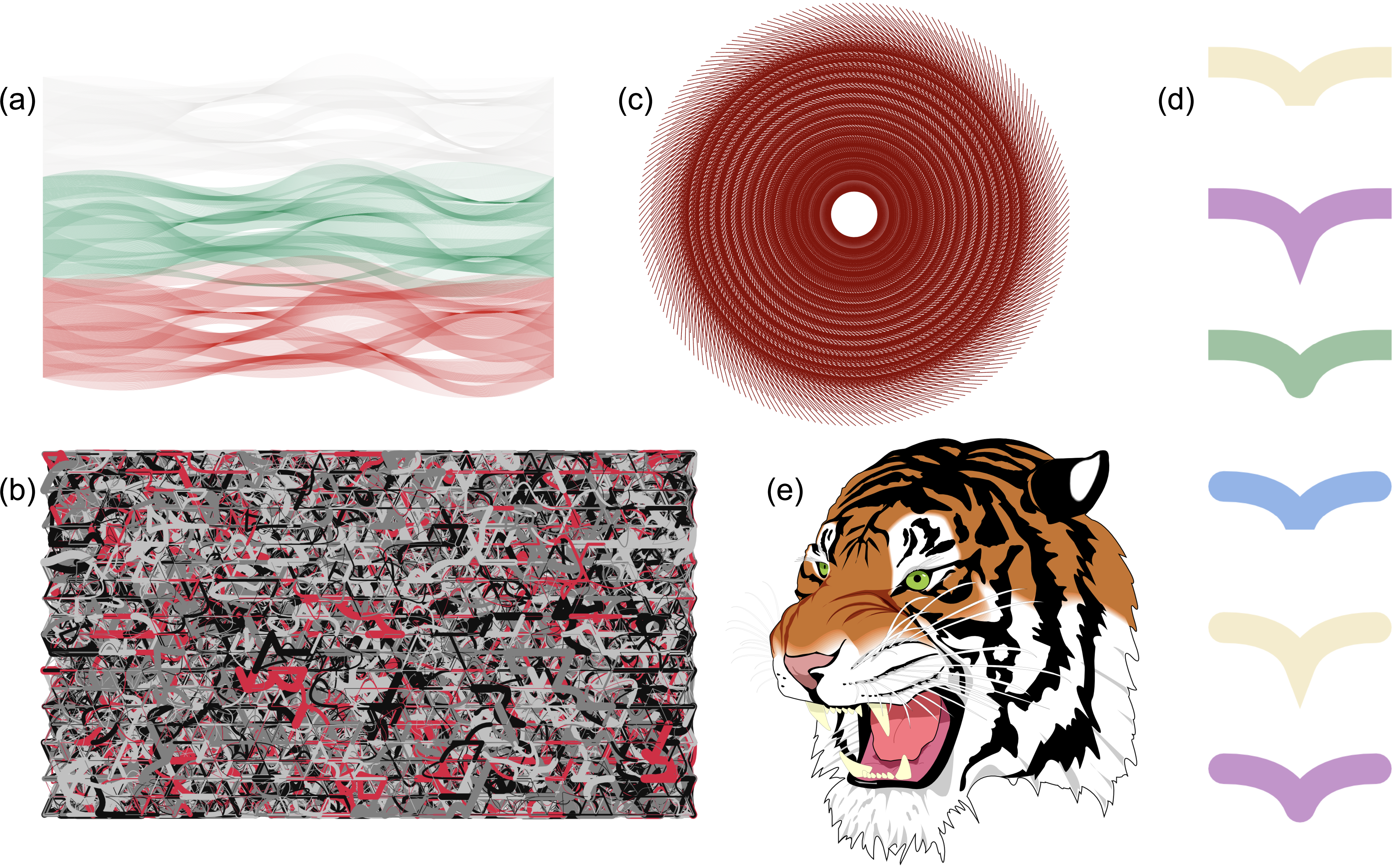}
    \caption{Renderings of some of our test scenes created using our algorithm. All images were rendered on an Apple M1 Max at 2088x1600 resolution using our GPU pipeline. \textbf{(a)} \emph{waves.svg} from the Nehab2020 timings data set (3.09~ms); \textbf{(b)} The \emph{mmark-120k} scene with 120,000 input segments (19.86~ms); \textbf{(c)} The \emph{long path} scene with \textasciitilde500k input segments and round caps (25~ms including CPU-side dashing -- the GPU time to render the scene is 9~ms); \textbf{(d)}; A simple test scene showcasing various cap and join styles (1.56~ms); \textbf{(e)} The famous \emph{Ghostscript Tiger} containing both stroked and filled paths (1.69~ms).}
    \Description{Renderings of the test scenes}
    \label{fig:renderings}
\end{figure*}

We instrumented our pipeline to obtain fine-grained pipeline stage execution times using the GPU \emph{timestamp query} feature supported by both Metal and Vulkan. This allowed us to collect measurements for the curve flattening and input processing stages in isolation. We ran the compute pipeline on the \emph{timings} SVG files provided by \citet{Nehab2020}. The largest of these SVGs is \emph{waves} (see rendering \emph{(a)} in Figure \ref{fig:renderings}) which contains 42 stroked paths and a total of 13,308 path segments.

We authored two additional test scenes in order to stress the GPU with a higher workload. The first is a very large stroked path with \textasciitilde500,000 individually styled dashed segments which we adapted from Skia's open-source test corpus. We used two variants of this scene with \emph{butt} and \emph{round} cap styles. For the second scene we adapted the \emph{Canvas Paths} test case from MotionMark~[\citenum{MotionMark}], which renders a large number of randomly generated stroked line, quadratic, and cubic Bézier paths. We made two versions with different sizes: mmark-70k and mmark-120k with 70,000 and 120,000 path segments, respectively. Table~\ref{table:bump-counts} shows the payload sizes of our largest test scenes.

\begin{table}
    \caption{Comparison of input and output segment counts in the more intensive GPU test cases. \emph{Segments} are input path segments. \emph{Lines} and \emph{Arcs} are the output primitives.}
    \begin{tabular}{|l|c|c|c|}
    \hline
        \multicolumn{1}{|c|}{\textbf{Test}} & \textbf{Input Segments} & \textbf{Lines} & \textbf{Arcs} \\
    \hline
        waves.svg & 13,308 & 475,855 & 181,229 \\
        mmark-70k & 70,000 & 1,577,705 & 1,162,117 \\
        mmark-120k & 120,000 & 2,709,013 & 1,994,946 \\
        long dash (butt) & 503,304 & 1,672,561 & 1,672,561 \\
        long dash (round) & 503,304 & 2,625,300 & 1,672,561 \\
    \hline
    \end{tabular}
    \label{table:bump-counts}
\end{table}

The test scenes were rendered to a GPU texture with dimensions 2088x1600. The contents of each scene were uniformly scaled up (by encoding a transform) to fit the size of the texture, in order to increase the required number of output primitives. As with the CPU timings, we used an error tolerance of 0.25 pixels in all cases. We recorded the execution times across 3,000 GPU submissions for each scene.

\begin{figure}
    \includegraphics[scale=0.69]{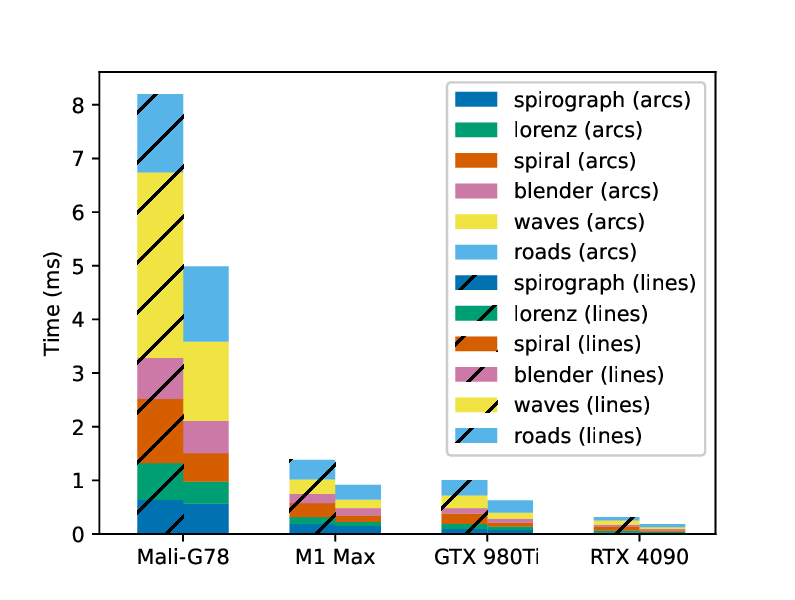}
    \caption{Mean GPU execution times of the compute shader that performs stroke expansion and flattening when run on the \citet{Nehab2020} \emph{timings} data set. The timings are shown in milliseconds. The hatched bars show the timings for the version that outputs line primitives while the solid bars show that for arcs. Mali-G78 (the lowest-end GPU we tested) converts \emph{waves} (the largest scene in the set) to lines in \textasciitilde3.5 ms. The compute shader executes at least an order of magnitude faster on the other GPUs.}
    \Description{A stacked bar chart of the GPU timings for stroke expansion from \citet{Nehab2020} timings SVGs}
    \label{fig:nehab-gpu-timings}
\end{figure}

\begin{figure}
    \includegraphics[scale=0.73]{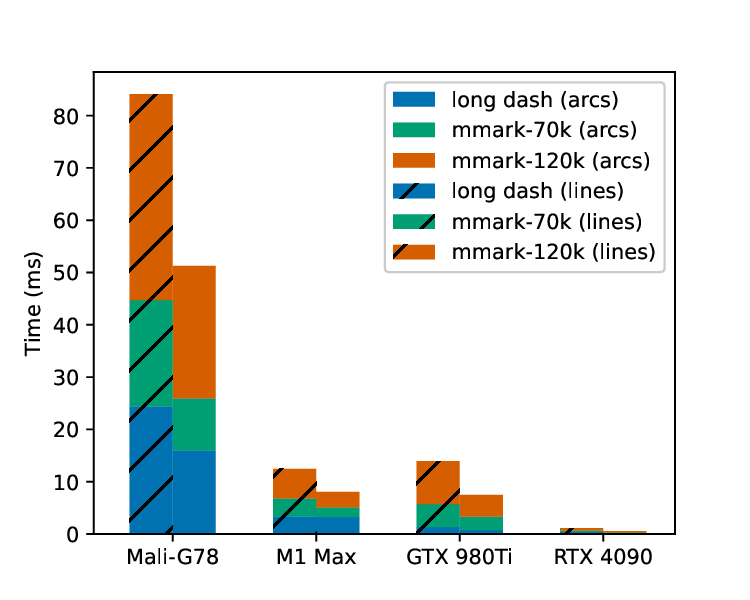}
    \caption{Mean GPU execution times of the compute shader that performs stroke expansion and flattening when run on the stress test scenes. The timings are shown in milliseconds.}
    \Description{A stacked bar chart of the GPU timings for stroke expansion from our test scenes}
    \label{fig:test-scenes-gpu-timings}
\end{figure}

Figures~\ref{fig:nehab-gpu-timings} and \ref{fig:test-scenes-gpu-timings} show the execution times for the two data sets. The entire \citet{Nehab2020} timings set adds up to less than 1.5~ms on all GPUs except the mobile unit. Our algorithm is \textasciitilde14$\times$ slower on the Mali-G78 compared to the M1 Max but it is still capable of processing \emph{waves} in 3.48~ms on average. We observed that the performance of the kernel scales well with increasing GPU ability even at very large workloads, as demonstrated by the order-of-magnitude difference in run time between the mobile, laptop, and high-end desktop form factors we tested. We also confirmed that outputting arcs instead of lines can lead to a 2$\times$ decrease in execution time (particularly in scenes with high curve content, like \emph{waves} and \emph{mmark}) as predicted by the lower overall primitive count.

\subsection{GPU: Input Processing} \label{subsection:encoding-results}

\begin{figure}
    \includegraphics[scale=0.66]{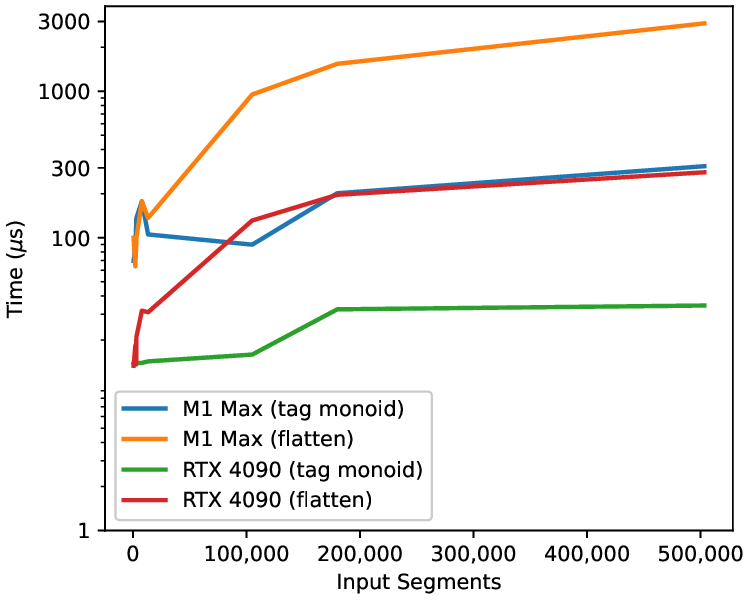}
    \caption{Mean GPU execution time of the tag monoid parallel prefix sums with increasing number of input path segments, measured on Apple M1 Max and NVIDIA RTX 4090. The figure also shows the execution time of the stroke expansion and flattening kernel. The y-axis is shown in logarithmic scale to highlight the similar scaling trend on both GPUs despite the order-of-magnitude difference in performance.}
    \Description{Mean GPU execution time of the tag monoid kernel}
    \label{fig:tag-monoid-gpu-timings}
\end{figure}

Figure~\ref{fig:tag-monoid-gpu-timings} shows the run time of the tag monoid parallel prefix sums with increasing input size. The prefix sums scale better at large input sizes compared to our stroke expansion kernel. This can be attributed to a lack of expensive computations and high control-flow uniformity in the tag monoid kernel. The CPU encoding times (prior to the GPU prefix sums) for some of the large scenes is shown in Table~\ref{table:bump-estimate-cost}. Our implementation applies dashing on the CPU. It is conceptually straightforward to implement on GPU; the core algorithm is prefix sum applied to segment length, and the Euler spiral representation is particularly well suited. However, we believe the design of an efficient GPU pipeline architecture involves comples tradeoffs which we did not sufficiently explore for this paper. Therefore the dash style applied to the \emph{long dash} scene was processed on the CPU and dashes were sequentially encoded as individual strokes. Including CPU-side dashing, the time to encode \emph{long dash} on the Apple M1 Max was approximately 25~ms on average, while the average time to encode the path after dashing was 4.14~ms.

Our GPU implementation targets graphics APIs that do not support dynamic memory allocation in shader code (in contrast to CUDA, which provides the equivalent of \emph{malloc}). This makes it necessary to allocate the storage buffer in advance with enough memory to store the output, however, the precise memory requirement is unknown prior to running the shader.

We augmented our encoding logic to compute a conservative estimate of the required output buffer size based on an upper bound on the number of line primitives per input path segment. We found that Wang's formula~[\citenum{Goldman2003}] works well in practice, as it is relatively inexpensive to evaluate, though it results in a higher memory estimate than required (see Table~\ref{table:bump-estimate-mem}). For parallel curves, Wang's formula does not guarantee an accurate upper bound on the required subdivisions, which we worked around by inflating the estimate for the source segment by a scale factor derived from the linewidth. We observed that enabling estimation increases our CPU pre-processing time by 2--3$\times$ on the largest scenes but overall the impact is negligible when the scene size is modest.

\begin{table}
    \caption{Comparison of CPU encoding time with and without size estimation on the largest scenes. All measurements were computed on an Apple M1 Max. The additional cost is negligible when the scene size is modest but tangible with increased complexity.}
    \begin{tabular}{|l|c|c|}
    \hline
        \multicolumn{1}{|c|}{\textbf{Test}} & \textbf{Encoding} & \textbf{Encoding + Estimation} \\
    \hline
        roads.svg  & 398.56 $\mu$s & 618.24 $\mu$s \\
        waves.svg  & 197.53 $\mu$s & 418.96 $\mu$s \\
        mmark-70k  & 2.69 ms       & 6.89 ms       \\
        mmark-120k & 4.28 ms       & 11.59 ms      \\
    \hline
    \end{tabular}
    \label{table:bump-estimate-cost}
\end{table}

\begin{table}
    \caption{The estimated and actual \emph{line soup} buffer sizes for scenes rendered at 2088x1600 resolution. Our estimation uses Wang's formula combined with a scaling heuristic based on the linewidth. The estimate is conservative and always overshoots the actual memory requirement.}
    \begin{tabular}{|l|c|c|}
    \hline
        \multicolumn{1}{|c|}{\textbf{Test}} & Actual Size & Estimated Size \\
    \hline
        roads.svg  & 2.76 MB   & 8.35 MB    \\
        waves.svg  & 10.89 MB  & 80.69 MB   \\
        mmark-70k  & 36.10 MB  & 103.06 MB  \\
        mmark-120k & 60.01 MB  & 177.33 MB  \\
    \hline
    \end{tabular}
    \label{table:bump-estimate-mem}
\end{table}

\section{Conclusions}

Path stroking has received attention in recent years, with \citet{Nehab2020} and \citet{Kilgard2020} both having proposed a complete theory of correct stroking in vector graphics. While there are a number of implementations of path filling on the GPU, the goal of also implementing stroke expansion on the GPU had not yet been realized. Building upon the theory proposed by \citet{Nehab2020}, we implemented an approach to stroke expansion that is GPU-friendly, and achieves high performance on commodity hardware.

We present the Euler spiral as an intermediate representation for a fast and precise approximation of both filled and stroked Bézier paths. Our method for lowering to both line and arc primitives avoids recursion and minimizes divergence, potentially serious problems for parallel evaluation on the GPU.  We propose a novel error metric to directly estimate approximation error as opposed to the commonly employed cut-then-measure approaches which often consume the lion's share of computational expense.

Our approach includes an efficient encoding scheme that alleviates the need for expensive CPU pre-computations and unlocks fully GPU-driven rendering of the entire vector graphics model.

\section{Future work}

\begin{itemize}
    \item The GPU implementation sequentializes some work that could be parallel, and is thus not as well load-balanced as it might be. An appealing future direction is to split the pipeline into separate stages, executed by the GPU as nodes in a work graph~[\citenum{Patel2024}]. This structure is appealing because load balancing is done by hardware.

    \item The pre-allocation requirement for bump-allocated line soup buffer is a limitation due to today's graphics APIs. A more accurate and optimized buffer size estimation heuristic is considered future work. It may also be interesting to explore whether graphics APIs can be extended to facilitate GPU-driven scheduling of the compute workload under a bounded memory footprint.

    \item Many of the error metrics were empirically determined. The mathematical theory behind them should be developed more rigorously, and that process will likely uncover opportunities to fine-tune the technique.

    \item As mentioned earlier, we did not have time to sufficiently explore a pipeline architecture for GPU-based dashing. Given its parameterization by arc length, we believe the Euler spiral representation lends itself well to an approach based on parallel prefix sums of segment arc lengths.
\end{itemize}


\bibliographystyle{ACM-Reference-Format}
\setcitestyle{numbers} 
\bibliography{paper}


\begin{thebibliography}{27}


\ifx \showCODEN    \undefined \def \showCODEN     #1{\unskip}     \fi
\ifx \showDOI      \undefined \def \showDOI       #1{#1}\fi
\ifx \showISBNx    \undefined \def \showISBNx     #1{\unskip}     \fi
\ifx \showISBNxiii \undefined \def \showISBNxiii  #1{\unskip}     \fi
\ifx \showISSN     \undefined \def \showISSN      #1{\unskip}     \fi
\ifx \showLCCN     \undefined \def \showLCCN      #1{\unskip}     \fi
\ifx \shownote     \undefined \def \shownote      #1{#1}          \fi
\ifx \showarticletitle \undefined \def \showarticletitle #1{#1}   \fi
\ifx \showURL      \undefined \def \showURL       {\relax}        \fi
\providecommand\bibfield[2]{#2}
\providecommand\bibinfo[2]{#2}
\providecommand\natexlab[1]{#1}
\providecommand\showeprint[2][]{arXiv:#2}

\bibitem[Mot(2021)]%
        {MotionMark}
 \bibinfo{year}{2021}\natexlab{}.
\newblock \bibinfo{title}{MotionMark 1.2}.
\newblock
\newblock
\urldef\tempurl%
\url{https://browserbench.org/MotionMark1.2/about.html}
\showURL{%
\tempurl}


\bibitem[Web(2024)]%
        {WebGPU}
World Wide Web Consortium \bibinfo{year}{2024}\natexlab{}.
\newblock \bibinfo{booktitle}{\emph{WebGPU}}.
\newblock World Wide Web Consortium.
\newblock
\urldef\tempurl%
\url{https://www.w3.org/TR/webgpu}
\showURL{%
\tempurl}


\bibitem[Connor and Krivodonova(2014)]%
        {Connor2014}
\bibfield{author}{\bibinfo{person}{Dale Connor} {and} \bibinfo{person}{Lilia
  Krivodonova}.} \bibinfo{year}{2014}\natexlab{}.
\newblock \showarticletitle{Interpolation of two-dimensional curves with Euler
  spirals}.
\newblock \bibinfo{journal}{\emph{J. Comput. Appl. Math.}}
  \bibinfo{volume}{261} (\bibinfo{year}{2014}), \bibinfo{pages}{320--332}.
\newblock
\showISSN{0377-0427}
\urldef\tempurl%
\url{https://doi.org/10.1016/j.cam.2013.11.009}
\showDOI{\tempurl}


\bibitem[Farouki and Neff(1990)]%
        {Farouki1990}
\bibfield{author}{\bibinfo{person}{R.~T. Farouki} {and} \bibinfo{person}{C.~A.
  Neff}.} \bibinfo{year}{1990}\natexlab{}.
\newblock \showarticletitle{Algebraic properties of plane offset curves}.
\newblock \bibinfo{journal}{\emph{Comput. Aided Geom. Des.}}
  \bibinfo{volume}{7}, \bibinfo{number}{1–4} (\bibinfo{date}{jun}
  \bibinfo{year}{1990}), \bibinfo{pages}{101–127}.
\newblock
\showISSN{0167-8396}
\urldef\tempurl%
\url{https://doi.org/10.1016/0167-8396(90)90024-L}
\showDOI{\tempurl}


\bibitem[Ganacim et~al\mbox{.}(2014)]%
        {Ganacim2014}
\bibfield{author}{\bibinfo{person}{Francisco Ganacim},
  \bibinfo{person}{Rodolfo~S. Lima}, \bibinfo{person}{Luiz~Henrique de
  Figueiredo}, {and} \bibinfo{person}{Diego Nehab}.}
  \bibinfo{year}{2014}\natexlab{}.
\newblock \showarticletitle{Massively-parallel vector graphics}.
\newblock \bibinfo{journal}{\emph{ACM Transactions on Graphics}}
  \bibinfo{volume}{33}, \bibinfo{number}{6} (\bibinfo{year}{2014}),
  \bibinfo{pages}{1--14}.
\newblock
\urldef\tempurl%
\url{https://doi.org/10.1145/2661229.2661274}
\showDOI{\tempurl}


\bibitem[gfx-rs authors(2024)]%
        {wgpu}
\bibfield{author}{\bibinfo{person}{The gfx-rs authors}.}
  \bibinfo{year}{2024}\natexlab{}.
\newblock \bibinfo{title}{gfx-rs/wgpu}.
\newblock
\newblock
\urldef\tempurl%
\url{https://github.com/gfx-rs/wgpu}
\showURL{%
\tempurl}


\bibitem[Goldman(2003)]%
        {Goldman2003}
\bibfield{author}{\bibinfo{person}{Ron Goldman}.}
  \bibinfo{year}{2003}\natexlab{}.
\newblock \showarticletitle{Chapter 5 - Bezier Approximation and Pascal's
  Triangle}.
\newblock In \bibinfo{booktitle}{\emph{Pyramid Algorithms}},
  \bibfield{editor}{\bibinfo{person}{Ron Goldman}} (Ed.).
  \bibinfo{publisher}{Morgan Kaufmann}, \bibinfo{address}{San Francisco},
  \bibinfo{pages}{187--306}.
\newblock
\showISBNx{978-1-55860-354-7}
\urldef\tempurl%
\url{https://doi.org/10.1016/B978-155860354-7/50006-4}
\showDOI{\tempurl}


\bibitem[Google(2024)]%
        {Skia}
\bibfield{author}{\bibinfo{person}{Google}.} \bibinfo{year}{2024}\natexlab{}.
\newblock \bibinfo{title}{Skia}.
\newblock
\newblock
\urldef\tempurl%
\url{https://skia.org}
\showURL{%
\tempurl}


\bibitem[Inc(2024)]%
        {Rive}
\bibfield{author}{\bibinfo{person}{Rive Inc}.} \bibinfo{year}{2024}\natexlab{}.
\newblock \bibinfo{title}{Rive Renderer}.
\newblock
\newblock
\urldef\tempurl%
\url{https://github.com/rive-app/rive-renderer}
\showURL{%
\tempurl}


\bibitem[Incorporated(2008)]%
        {PDF2008}
\bibfield{author}{\bibinfo{person}{Adobe~Systems Incorporated}.}
  \bibinfo{year}{2008}\natexlab{}.
\newblock \bibinfo{title}{Document management -- Portable document format --
  Part 1: PDF 1.7}.
\newblock
\newblock
\urldef\tempurl%
\url{https://opensource.adobe.com/dc-acrobat-sdk-docs/pdfstandards/PDF32000_2008.pdf}
\showURL{%
\tempurl}


\bibitem[Kilgard(2020a)]%
        {Kilgard2020a}
\bibfield{author}{\bibinfo{person}{Mark~J. Kilgard}.}
  \bibinfo{year}{2020}\natexlab{a}.
\newblock \bibinfo{title}{Anecdotal Survey of Variations in Path Stroking among
  Real-world Implementations}.
\newblock
\newblock
\showeprint[arxiv]{2007.12254}


\bibitem[Kilgard(2020b)]%
        {Kilgard2020}
\bibfield{author}{\bibinfo{person}{Mark~J. Kilgard}.}
  \bibinfo{year}{2020}\natexlab{b}.
\newblock \showarticletitle{Polar Stroking: New Theory and Methods for Stroking
  Paths}.
\newblock \bibinfo{journal}{\emph{ACM Trans. Graph.}} \bibinfo{volume}{39},
  \bibinfo{number}{4}, Article \bibinfo{articleno}{145} (\bibinfo{date}{Aug.}
  \bibinfo{year}{2020}), \bibinfo{numpages}{15}~pages.
\newblock
\showISSN{0730-0301}
\urldef\tempurl%
\url{https://doi.org/10.1145/3386569.3392458}
\showDOI{\tempurl}


\bibitem[Kimia et~al\mbox{.}(2003)]%
        {Kimia2003}
\bibfield{author}{\bibinfo{person}{Benjamin~B. Kimia}, \bibinfo{person}{Ilana
  Frankel}, {and} \bibinfo{person}{Ana-Maria Popescu}.}
  \bibinfo{year}{2003}\natexlab{}.
\newblock \showarticletitle{Euler Spiral for Shape Completion}.
\newblock \bibinfo{journal}{\emph{International Journal of Computer Vision}}
  \bibinfo{volume}{54}, \bibinfo{number}{1} (\bibinfo{date}{01 Aug}
  \bibinfo{year}{2003}), \bibinfo{pages}{159--182}.
\newblock
\showISSN{1573-1405}
\urldef\tempurl%
\url{https://doi.org/10.1023/A:1023713602895}
\showDOI{\tempurl}


\bibitem[Laine and Karras(2011)]%
        {Laine2011}
\bibfield{author}{\bibinfo{person}{Samuli Laine} {and} \bibinfo{person}{Tero
  Karras}.} \bibinfo{year}{2011}\natexlab{}.
\newblock \showarticletitle{High-Performance Software Rasterization on GPUs}.
\newblock \bibinfo{journal}{\emph{HPG '11: Proceedings of the ACM SIGGRAPH
  Symposium on High Performance Graphics}} (\bibinfo{year}{2011}),
  \bibinfo{pages}{79--88}.
\newblock
\urldef\tempurl%
\url{https://doi.org/10.1145/2018323.2018337}
\showDOI{\tempurl}


\bibitem[Levien(2021)]%
        {Levien2021}
\bibfield{author}{\bibinfo{person}{Raph Levien}.}
  \bibinfo{year}{2021}\natexlab{}.
\newblock \bibinfo{title}{Cleaner parallel curves with Euler spirals}.
\newblock
\newblock
\urldef\tempurl%
\url{https://raphlinus.github.io/curves/2021/02/19/parallel-curves.html}
\showURL{%
\tempurl}


\bibitem[Maier(2014)]%
        {Maier2014}
\bibfield{author}{\bibinfo{person}{Georg Maier}.}
  \bibinfo{year}{2014}\natexlab{}.
\newblock \showarticletitle{Optimal arc spline approximation}.
\newblock \bibinfo{journal}{\emph{Computer Aided Geometric Design}}
  \bibinfo{volume}{31}, \bibinfo{number}{5} (\bibinfo{year}{2014}),
  \bibinfo{pages}{211--226}.
\newblock
\urldef\tempurl%
\url{https://doi.org/10.1016/j.cagd.2014.02.011}
\showDOI{\tempurl}


\bibitem[Meek and Walton(2004)]%
        {Meek2004}
\bibfield{author}{\bibinfo{person}{D.~S. Meek} {and} \bibinfo{person}{D.~J.
  Walton}.} \bibinfo{year}{2004}\natexlab{}.
\newblock \showarticletitle{An arc spline approximation to a clothoid}.
\newblock \bibinfo{journal}{\emph{J. Comput. Appl. Math.}}
  \bibinfo{volume}{170}, \bibinfo{number}{1} (\bibinfo{year}{2004}),
  \bibinfo{pages}{59--77}.
\newblock
\urldef\tempurl%
\url{https://doi.org/10.1016/j.cam.2003.12.038}
\showDOI{\tempurl}


\bibitem[Narayan(2014)]%
        {Narayan2014}
\bibfield{author}{\bibinfo{person}{Smita Narayan}.}
  \bibinfo{year}{2014}\natexlab{}.
\newblock \showarticletitle{Approximating Cornu spirals by arc splines}.
\newblock \bibinfo{journal}{\emph{J. Comput. Appl. Math.}}
  \bibinfo{volume}{255}, \bibinfo{number}{1} (\bibinfo{year}{2014}).
\newblock
\urldef\tempurl%
\url{https://doi.org/10.1016/j.cam.2013.06.038}
\showDOI{\tempurl}


\bibitem[Nehab(2020)]%
        {Nehab2020}
\bibfield{author}{\bibinfo{person}{Diego Nehab}.}
  \bibinfo{year}{2020}\natexlab{}.
\newblock \showarticletitle{Converting Stroked Primitives to Filled
  Primitives}.
\newblock \bibinfo{journal}{\emph{ACM Trans. Graph.}} \bibinfo{volume}{39},
  \bibinfo{number}{4}, Article \bibinfo{articleno}{137} (\bibinfo{date}{Aug.}
  \bibinfo{year}{2020}), \bibinfo{numpages}{17}~pages.
\newblock
\showISSN{0730-0301}
\urldef\tempurl%
\url{https://doi.org/10.1145/3386569.3392392}
\showDOI{\tempurl}


\bibitem[Nuntawisuttiwong and Dejdumrong(2021)]%
        {Nuntawisuttiwong2021}
\bibfield{author}{\bibinfo{person}{Taweechai Nuntawisuttiwong} {and}
  \bibinfo{person}{Natasha Dejdumrong}.} \bibinfo{year}{2021}\natexlab{}.
\newblock \showarticletitle{An Approximation of Bézier Curves by a Sequence of
  Circular Arcs}.
\newblock \bibinfo{journal}{\emph{Information Technology and Control}}
  \bibinfo{volume}{50}, \bibinfo{number}{2} (\bibinfo{year}{2021}).
\newblock
\urldef\tempurl%
\url{https://doi.org/10.5755/j01.itc.50.2.25178}
\showDOI{\tempurl}


\bibitem[Patel and Riddell(2024)]%
        {Patel2024}
\bibfield{author}{\bibinfo{person}{Amar Patel} {and} \bibinfo{person}{Tex
  Riddell}.} \bibinfo{year}{2024}\natexlab{}.
\newblock \bibinfo{booktitle}{\emph{D3D12 Work Graphs}}.
\newblock DirectX Developer Blog.
\newblock
\urldef\tempurl%
\url{https://devblogs.microsoft.com/directx/d3d12-work-graphs/}
\showURL{%
\tempurl}


\bibitem[Reif and Weinmann(2021)]%
        {Reif2021}
\bibfield{author}{\bibinfo{person}{Ulrich Reif} {and} \bibinfo{person}{Andreas
  Weinmann}.} \bibinfo{year}{2021}\natexlab{}.
\newblock \showarticletitle{Clothoid fitting and geometric Hermite
  subdivision}.
\newblock \bibinfo{journal}{\emph{Advances in Computational Mathematics}}
  \bibinfo{volume}{47}, \bibinfo{number}{50} (\bibinfo{date}{26 June}
  \bibinfo{year}{2021}).
\newblock
\urldef\tempurl%
\url{https://doi.org/10.1007/s10444-021-09876-5}
\showDOI{\tempurl}


\bibitem[Tiller and Hanson(1984)]%
        {Tiller1984}
\bibfield{author}{\bibinfo{person}{W. Tiller} {and} \bibinfo{person}{E.~G.
  Hanson}.} \bibinfo{year}{1984}\natexlab{}.
\newblock \showarticletitle{Offsets of two-dimensional profiles}.
\newblock \bibinfo{journal}{\emph{IEEE Computer Graphics and Applications}}
  \bibinfo{volume}{4}, \bibinfo{number}{9} (\bibinfo{date}{Sept.}
  \bibinfo{year}{1984}), \bibinfo{pages}{36--46}.
\newblock


\bibitem[Walton and Meek(2009)]%
        {Walton2009}
\bibfield{author}{\bibinfo{person}{D.~J. Walton} {and} \bibinfo{person}{D.~S.
  Meek}.} \bibinfo{year}{2009}\natexlab{}.
\newblock \showarticletitle{G1 interpolation with a single Cornu spiral
  segment}.
\newblock \bibinfo{journal}{\emph{J. Comput. Appl. Math.}}
  \bibinfo{volume}{223}, \bibinfo{number}{1} (\bibinfo{year}{2009}),
  \bibinfo{pages}{86--96}.
\newblock
\showISSN{0377-0427}
\urldef\tempurl%
\url{https://doi.org/10.1016/j.cam.2007.12.022}
\showDOI{\tempurl}


\bibitem[Wieleitner(1907)]%
        {Wieleitner1907}
\bibfield{author}{\bibinfo{person}{Heinrich Wieleitner}.}
  \bibinfo{year}{1907}\natexlab{}.
\newblock \showarticletitle{Die Parallelkurve der Klothoide}.
\newblock \bibinfo{journal}{\emph{Archiv der Mathematik und Physik}}
  \bibinfo{volume}{11} (\bibinfo{year}{1907}), \bibinfo{pages}{373--375}.
\newblock


\bibitem[Yoshida and Saito(2012)]%
        {Yoshida2012}
\bibfield{author}{\bibinfo{person}{Norimasa Yoshida} {and}
  \bibinfo{person}{Takafumi Saito}.} \bibinfo{year}{2012}\natexlab{}.
\newblock \showarticletitle{The Evolutes of Log-Aesthetic Planar Curves and the
  Drawable Boundaries of the Curve Segments}.
\newblock \bibinfo{journal}{\emph{Computer-Aided Design and Applications}}
  \bibinfo{volume}{9}, \bibinfo{number}{5} (\bibinfo{year}{2012}),
  \bibinfo{pages}{721--731}.
\newblock
\urldef\tempurl%
\url{https://doi.org/10.3722/cadaps.2012.721-731}
\showDOI{\tempurl}


\bibitem[Yzerman(2020)]%
        {Yzerman2020}
\bibfield{author}{\bibinfo{person}{Fabian Yzerman}.}
  \bibinfo{year}{2020}\natexlab{}.
\newblock \bibinfo{title}{Fast approaches to simplify and offset Bézier curves
  within specified error limits}.
\newblock
\newblock
\urldef\tempurl%
\url{https://blend2d.com/research/simplify_and_offset_bezier_curves.pdf}
\showURL{%
\tempurl}


\end{thebibliography}

\appendix

\section{Geometric Hermite interpolation for Euler spiral}
\label{appendix:gh}

This appendix gives the detailed algorithm for Geometric Hermite interpolation, determining Euler spiral segment parameters given the tangent angles at the endpoints relative to the chord.

The Euler spiral is represented by $\kappa(s) = k_0 + k_1s$, where $s$ ranges from 0 to 1, i.e., the arc length is unit normalized. This segment is scaled, rotated, and translated into place so that its endpoints match the desired locations. Immediately, $k_0$ can be determined as $\theta_0 + \theta_1$. Therefore, the remaining parameters to compute are $k_1$ and the ratio of chord length to arc length. The latter \emph{could} be computed from $k_0$ and $k_1$ by numerical integration, but it is more efficient to obtain it at the same time as $k_1$. Setting $k = \theta_0 + \theta_1$ and $\Delta = \theta_1 - \theta_0$, we have:

\[
    \setlength\arraycolsep{2pt}
    \begin{array}{rl}
    k_1 = & 6\Delta \\
     - & \Delta^3 / 70 \\
     - & \Delta^5 / 10780 \\
     + & \Delta^7 \cdot 2.769178184818219\times 10^{-7} \\
     - & k^2\Delta / 10 \\
     + & k^2\Delta^3 / 4200 \\
     + & k^2\Delta^5 \cdot 1.6959677820260655\times 10^{-5} \\
     - & k^4\Delta / 1400 \\
     + & k^4\Delta^3 \cdot 6.84915970574303\times 10^{-5} \\
     - & k^6\Delta \cdot 7.936475029053326\times 10^{-6}
    \end{array}
\]

Similarly, the formula for the chord to arc length ratio:

\[
    \setlength\arraycolsep{2pt}
    \begin{array}{rl}
    c = & 1 \\
    - & \Delta^2 / 40 \\
    + & \Delta^4 \cdot 0.00034226190482569864 \\
    - & \Delta^6 \cdot 1.9349474568904524\times 10^{-6} \\
    - & k^2 / 24 \\
    + & k^2\Delta^2 \cdot 0.0024702380951963226 \\
    - & k^2\Delta^4 \cdot 3.7297408997537985\times 10^{-5} \\
    + & k^4 / 1920 \\
    - & k^4\Delta^2 \cdot 4.87350869747975\times 10^{-5} \\
    - & k^6 \cdot 3.1001936068463107\times 10^{-6}
    \end{array}
\]

Note that the latter quantity is equal to $\mbox{sinc}(k/2)$ when $\Delta = 0$.

The coefficients were determined by numerical differentiation, using a Newton-style solver as the source of truth. The resulting formulas are extremely accurate over a wide range of inputs.


\end{document}